# Mechanisms of Calcium Leak from Cardiac Sarcoplasmic Reticulum Revealed by Statistical Mechanics

**Running title: Classification of calcium leak in heart**

Anna V. Maltsev[1], Michael D. Stern[2], Victor A. Maltsev[2]*

[1]School of Mathematics, Queen Mary University of London, London, United Kingdom

[2]Laboratory of Cardiovascular Science, Biomedical Research Center, Intramural Research Program, National Institute on Aging, NIH, Baltimore, Maryland, USA.

***Corresponding author:**

Victor A. Maltsev, PhD
Email: MaltsevVi@mail.nih.gov
Telephone: 410-558-8426
Fax: 410-558-8150

**Abstract**

Heart muscle contraction is normally activated by a synchronized Ca release from sarcoplasmic reticulum (SR), a major intracellular Ca store. However, under abnormal conditions Ca leaks from the SR, decreasing heart contraction amplitude and increasing risk of life-threatening arrhythmia. The mechanisms and regimes of SR operation generating the abnormal Ca leak remain unclear. Here we employed both numerical and analytical modeling to get mechanistic insights into the emergent Ca leak phenomenon. Our numerical simulations using a detailed realistic model of Ca release unit (CRU) reveal sharp transitions resulting in Ca leak. The emergence of leak is closely mapped mathematically to the Ising model from statistical mechanics. The system steady-state behavior is determined by two aggregate parameters: the analogues of magnetic field ($h$) and the inverse temperature ($β$) in the Ising model, for which we have explicit formulas in terms of SR Ca and release channel opening/closing rates. The classification of leak regimes takes the shape of a phase $β$-$h$ diagram, with the regime boundaries occurring at $h=0$ and a critical value of $β$ ($β^*$) which we estimate using a classical Ising model and mean field theory. Our theory predicts that a synchronized Ca leak will occur when $h>0$ and $β>β^*$ and a disordered leak occurs when $β<β^*$ and $h$ is not too negative. The disorder leak is distinguished from synchronized leak (in long-lasting sparks) by larger Peierls contour lengths, an output parameter reflecting degree of disorder. Thus, in addition to our detailed numerical model approach we also offer an instantaneous computational tool using analytical formulas of the Ising model for respective RyR parameters and SR Ca load that describe and classify phase transitions and leak emergence.

**Statement of Significance**

This report provides new quantitative insight into problem of Ca leak from sarcoplasmic reticulum (SR). Our numerical model simulations discovered sharp transitions in Ca release unit operation resulting in Ca leak. The leak emergence is closely mapped mathematically to the Ising model from statistical mechanics, namely to two types of phase transitions known for this model: magnetization (i.e. spontaneous synchronization of spins' orientation) and the Onsager's order-disorder transition. Thus, our model offers a new classification of leak that takes the form of a phase diagram representing normal function and two leak types: disorder leak and synchronized leak. The model also offers an instantaneous computational tool to describe phase transitions as a function of release channel and SR parameters.

**Glossary**

| | |
|---|---|
| RyR | ryanodine receptor (Ca release channel) |
| SR | sarcoplasmic reticulum (a major intracellular Ca store in cardiac myocytes) |
| FSR | free SR (also known as network SR) |
| JSR | junctional SR |
| $Ca_{FSR}$ | [Ca] in FSR |
| $Ca_{JSR}$ | [Ca] in JSR |
| CRU | Ca release unit that includes JSR with RyRs and L-type Ca channels |
| CICR | Ca-induced Ca release |
| $Ca_{dyad}$ | local [Ca] in dyadic space |
| $h$ | isomorphic analog of magnetic field in our Ising model |
| $β$ | isomorphic analog of the inverse temperature in our Ising model. |
| $β*$ | critical $β$ at which CRU undergoes "order-disorder" transition |
| $k_o$ | RyR opening rate given as $k_o=λ*exp(γ*Ca_{dyad})$ |
| $C$ | RyR closing rate ($C=0.117$ ms$^{-1}$) |
| $P_o$ | the probability to find RyR in the open state |
| $i_{RyR}$ | unitary current via one RyR |
| $Ψ$ | interaction profile, defined as local [Ca] distribution in dyadic space caused by $i_{RyR}$ |
| $N_{RyR}$ | number of RyRs in a CRU |
| $n_∞$ | average number of open RyR at steady-state |
| $U$ | distance between neighboring RyRs in CRU ($U=30$ nm) |

**Introduction**

In heart muscle electrical excitation is coupled to contraction *via* Ca signaling between L-type Ca channels of the plasma membrane and Ca release channels (ryanodine receptors, RyRs) residing in the sarcoplasmic reticulum (SR), a major Ca store within cardiac cells. Opening of L-type Ca channels leads to synchronous openings of neighboring RyRs as the open probability of RyR is increased by cytoplasmic [Ca]. The resultant Ca-induced-Ca-release (CICR)(1) triggers displacement of myofilaments and cell contraction.

The synchronized activation of RyRs during systole is followed by the robust release termination causing muscle relaxation. However, under pathological conditions Ca release does not terminate but continues during diastole (2), causing a Ca leak. Deteriorating effects of Ca leak on heart function include (2,3): 1) reduced systolic SR Ca levels leading to systolic dysfunction; 2) elevated diastolic Ca leading to diastolic dysfunction; 3) energy drain to repump Ca; 4) triggered arrhythmias.

The mechanisms and regimes of SR operation generating abnormal leak remain elusive. The RyR is a huge molecule featuring extremely complex regulation *via* numerous post-translational modifications and multiple regulatory proteins (2,4,5). In pathological conditions Ca leak can be caused by increased sensitivity of RyRs to Ca due to RyR phosphorylation by CaMKII (2,3). Furthermore, diastolic Ca leak is a multiscale complex phenomenon. In ventricular myocytes RyRs are organized in clusters of 10-300 channels (6) residing in the junctional SR (JSR), forming Ca release units (CRU), which can generate a local elementary Ca release dubbed Ca spark (7,8). Ca leak can range from "invisible" or ''nonspark'' events originating from openings of just one or several RyRs (9), diastolic Ca sparks (8), macro-sparks, small abortive Ca waves, up to Ca waves of a cell size (review (10)). The invisible releases and sparks that fail to terminate have been extensively studied using computational models (11-17). While numerical simulations showed that the leak is facilitated by rapid inter- and intra-SR Ca diffusion (15,16) or faster JSR refilling rate with Ca (13), the mechanisms of the leak as an emerging, macroscopic phenomenon remain unclear.

Thus, our understanding of Ca leak would greatly benefit if the numerical modeling were combined with an appropriate statistical model to describe the collective behavior of RyRs underlying leak emergence. Regulation of cardiac muscle contraction strength has been described *via* statistics of success and failure of L-type Ca channels to ignite a Ca spark (7). More recently we also showed that the lattice of open and closed RyRs in CRU (Fig. S1) and lattice of spins in ferromagnets behave mathematically identically, namely as the Ising model from statistical physics (18). Our mapping CRU to a lattice of spins is based on a clear analogy of interactions between RyRs and spins. Indeed, spins could be in two positions + or -, and RyRs also can be either in open or closed state. A spin "wants" to turn its neighbor to same state, and RyRs do the same: an open RyR also "wants" to open its neighbor *via* CICR, whereas a closed RyR "wants" to close its neighbor *via* interrupting ongoing CICR (13) (i.e. induction decay (19) or "pernicious attrition" (20)). In our mapping RyRs become and act as spins and their Ca profiles become interaction profiles.

Using this approach we have shown that sparks normally terminate *via* a classical transition known for the lattice of spins as magnetization or polarity reversal, when magnetic field (*h*) changes sign from positive to negative (18). Magnetization in the classical Ising model is defined as the number of plus-spins minus the number of minus-spins, divided by the total number of spins to normalize it. It depends on the external magnetic field *h* and temperature. When the temperature is near zero, spin-to-spin interactions are very strong and the spins will align yielding a magnetization of plus or minus 1. The sign of the external magnetic field will determine whether it is plus or minus 1 that minimizes the configuration energy. As temperature increases, the spin interactions wane but the magnetization will stay near plus or minus 1. As the temperature approaches a critical value of the temperature parameter (known as Curie temperature), magnetization will decrease, hitting a first-order phase transition (Onsager's order-disorder transition) at Curie temperature. In terms of system behavior, as the temperature increases to above Curie temperature, the spins become effectively independent and the change from order to disorder is abrupt. Once this transition occurs, the value of magnetic field plays a key role in establishing what proportion of spins is up. As we have established in (18) that the CRU is isomorphic to an Ising model, all these phenomena are expected to occur also in the CRU, which is exactly the main purpose of this paper.

The order-disorder transition for the classical nearest-neighbor Ising model was first shown by Peierls in (22). The model we use here has interactions that are more than nearest neighbor interaction but have fast decay at infinity. The statement of the analogous order-disorder transition appears as Proposition 2.1 in (23) with proof and supporting references and history. In the present paper we use the term "phase transition" with respect to the analytical Ising model, but "phase-like transition", when we consider our numerical Ca spark model (13).

In our previous paper (18) from our exact mapping we have derived formulas for the analogues of the magnetic field $h$ and inverse temperature $\beta$ in terms of natural biophysical parameters:

(1) $$h = \frac{1}{2\beta} \ln\left(\frac{\lambda}{C}\right) + 2\pi \int_{r>.5} \phi(r, Ca_{JSR}) dr$$

(2) $$\beta = \gamma \psi(U, Ca_{JSR})/4$$

where $\phi(r) = \psi(Ur)/\psi(U)$ is a normalized and scaled version of $\psi(r)$ at each given $Ca_{JSR}$ and grid size. The function $\psi$ gives $Ca_{dyad}$ as a function of distance $r$ from one open RyR and $U$ is the distance to the channel nearest neighbor (Fig. S2A). Here the constants $\lambda$, $C$, and $\gamma$ are parameters of RyR channel kinetics, namely $C$ is the closing rate, while $\lambda$ and $\gamma$ correspond to parameters of an exponential in local Ca ($Ca_{dyad}$) that describes the RyR opening rate (Fig. S2C). Further details on the relationship between the abstract Ising model and ion channels and CRU parameters are given in Table S1. Derivation of Equations 1 and 2 are given in section "Ising model methods" in Supporting Materials and Methods.

Here we employed a realistic detailed numerical model of a Ca release unit (13) combined with our analytical Ising approach (18) to explore and classify leak regimes generated by the statistical ensemble of RyRs within a CRU for various values of $h$ and $\beta$. Thus, our sensitivity analyses of the numerical model were guided by predictions of the analytical model in terms of specific model parameters and their ranges for the simulations. Our study rationale and results are summarized in a respective phase diagram (Fig. 1), in which model behaviors are classified based on their values of $h$ and $\beta$ and their relation to particular values of extreme importance namely, $h = 0$ and $\beta = \beta^*$. We identified two leak regimes: one is linked to a failure of $h$ to change its sign and the other is linked to disorder, i.e. Onsager's "order-disorder" transition. In magnetism, disorder occurs when the temperature becomes too high (above Curie temperature), overwhelming the interactions. In the CRU model, when RyR interactions are weakened for example by a low SR level, we observe a similar disordered state. We derive analytical formulas for exact relations between RyR parameters and SR Ca load for both leak regimes. Both leak regimes were found in our numerical model simulations.

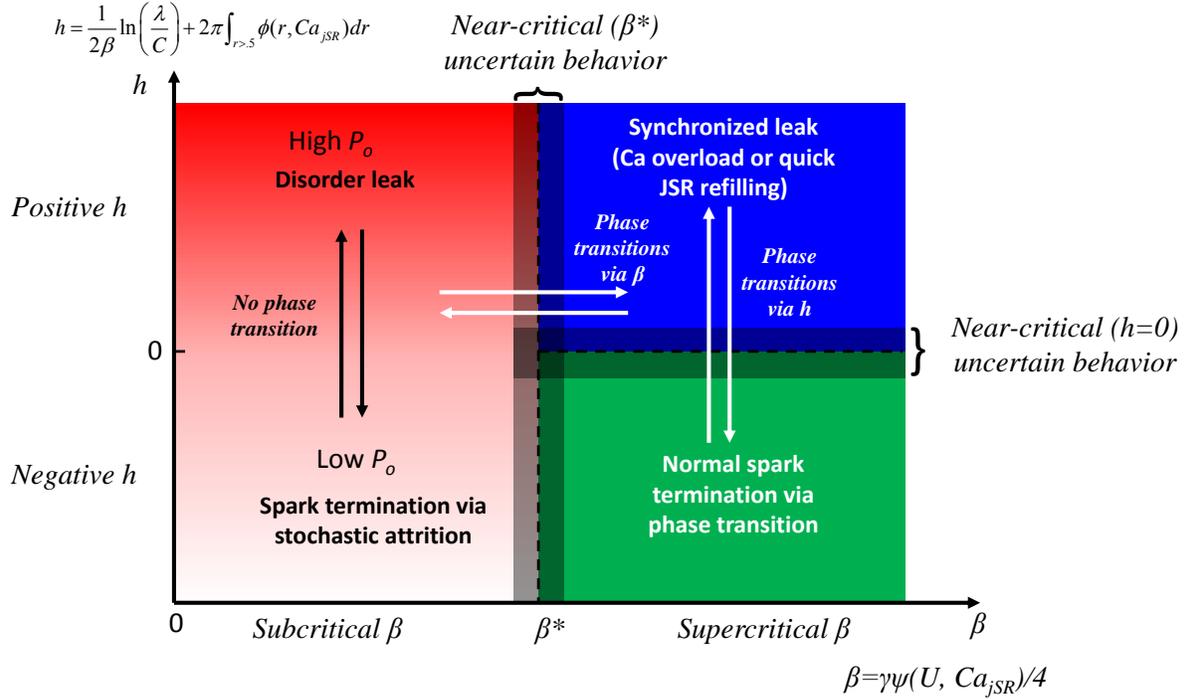

**Fig. 1**: **The phase diagram of CRU operation predicted by analytical Ising model.** Relations of spark termination regimes and types of Ca leaks with respect to phase transitions *via h* polarity reversal and *via β\**. The different regimes of CRU operation with phase transition boundaries are shown by different colors. In subcritical *β* regime red shade shows the degree of leak, which is gradual (i.e. not a phase transition).

**Methods**

We used our previously reported numerical model of a Ca spark (Stern model) (13). The model is illustrated in Fig. S1. We also used a recent analytical model of Ca spark termination *via* an Ising-type phase transition (18) (Supporting Materials and Methods). We constructed an exact mapping between two systems in 2 dimensions: (i) the lattice of interacting RyRs and (ii) the Ising model of interacting spins within a ferromagnet, so that both systems are described by the same mathematical formulations. We further formulated four general conditions for a system of RyRs to satisfy the mapping:

1. The RyRs are arranged in a lattice structure, with a nearest neighbor distance $U$.

2. Each RyR can be in an open or closed state, corresponding to the plus and minus states of the Ising model.

3. At each given [Ca] in JSR ($Ca_{JSR}$), the interaction profile $\psi(r, Ca_{JSR})$, defined as local [Ca] distribution in dyadic space caused by $i_{RyR}$ (Fig. S2A), is roughly stable in time (Fig. S2B), corresponding to time-invariance of RyR interactions, and roughly the same for any RyR in the lattice, corresponding to translation invariance. In the present study we demonstrated a minor effect of driving force reduction due to open neighboring RyRs on interaction profile (Fig. S3 and Table S2).

4. The RyR opening rate $k_o$ depends exponentially on $Ca_{dyad}$, i.e. $k_o = \lambda * exp(\gamma * Ca_{dyad})$ (Fig. S2C, red curve), and the closing rate $C$ is a constant ($C = 0.117$ ms$^{-1}$ in Stern model).

All four above conditions are met in our CRU model (18).

**Results**
*Estimates of β\* using analytical approaches*

The exact value of *β\** is known for the classical Ising model, in which spin interactions are limited to the nearest neighbors. However, in the Ising model that is isomorphic to the CRU, the RyRs (corresponding to spins) interact *via* a Gaussian-like interaction profile (Fig. S2A), i.e. the interactions spread to other, more distant neighbors within the RyR grid. While this precludes a precise analytic computation of *β\**, the interaction strength notably weakens for the second (and higher) order neighbors, so that the exact value of *β\** ($\beta^*_{Ising}$=0.138, see Supporting Materials and Methods) of classical model can serve as an upper-bound estimate for the "order-disorder" transition in the CRU. We also obtained a lower bound estimate of *β\** using the mean field approximation ($\beta^*_{mean\_field}$=0.0784, Supporting Materials and Methods) that represents an extreme case when all RyRs in a CRU would interact equally. Thus, our *β\** estimates from the two analytical approaches provide a fairly narrow range for the true *β\** in our CRU system, i.e. $0.0784 < \beta^* < 0.138$.

*Order-disorder phase-like transition and the value of critical β in numerical simulations*

We tested if "order-disorder" phase-like transition can occur *via β* at *h*=0 in a CRU by varying *λ* and *γ* in numerical simulations. We achieved *h*=0 by clamping $Ca_{JSR}$ at 0.1 mM, so that the integral in Equation 1 and interaction profile *ψ* in Equation 2 became fixed. Solving Equation 1 for *h*=0 yields a one-to-one correspondence between *β* and *λ* (with *C*=0.117 ms$^{-1}$ =const). As $Ca_{JSR}$ is fixed, so is *ψ(U)*, and hence Equation 2 yields a one-to-one correspondence between *β* and *γ*. Using this correspondence, we computed *β* and *λ* pairs for *γ* varying from 0.02 to 0.15μM$^{-1}$ (Table S3). The plots of opening rate ($k_o$) vs. local [Ca] in dyadic space ($Ca_{dyad}$) for respective pairs of *λ* and *γ* are shown in Fig. 2A. For each *β* and its respective *λ*-*γ* pair we performed a simulation of 100 sparks using Stern model (examples in Fig. 2B) and for each spark we determined its median extinction time, i.e. when all RyRs become closed. As *β* decreased, we observed a sharp increase in extinction times. In simulations with *β*>0.13 sparks robustly terminated, with median extinction times being roughly between 50 and 200 ms, but with *β*<0.1, the median extinction times suddenly became basically infinite on the heartbeat cycle scale, i.e.>10,000 ms (Fig. 2C). Thus, the phase-like transition in our CRU model occurred at *β\**~0.1, i.e. indeed between the analytically computed upper and lower bounds of *β\** (Fig. 2C, gray band).

Using electron microscopy, Franzini-Armstrong et al. reported average numbers of RyRs per couplon in dyads: 90, 128, and 267, for dog, mouse, and rat, respectively (Table 2 in (6)). Thus, Fig. 2C shows the phase-like transition for a relatively small CRU of 81 RyRs. Our additional numerical simulations revealed a similar phase-like transition for a much larger cluster of 169 RyRs (Fig. S4). Thus, our results indicate that the phase-like transitions in the numerical model of Ca spark occur independently of the number of channels in the CRU and their β* can be well-approximated analytically within the physiological range of CRU sizes known for mammalian species.

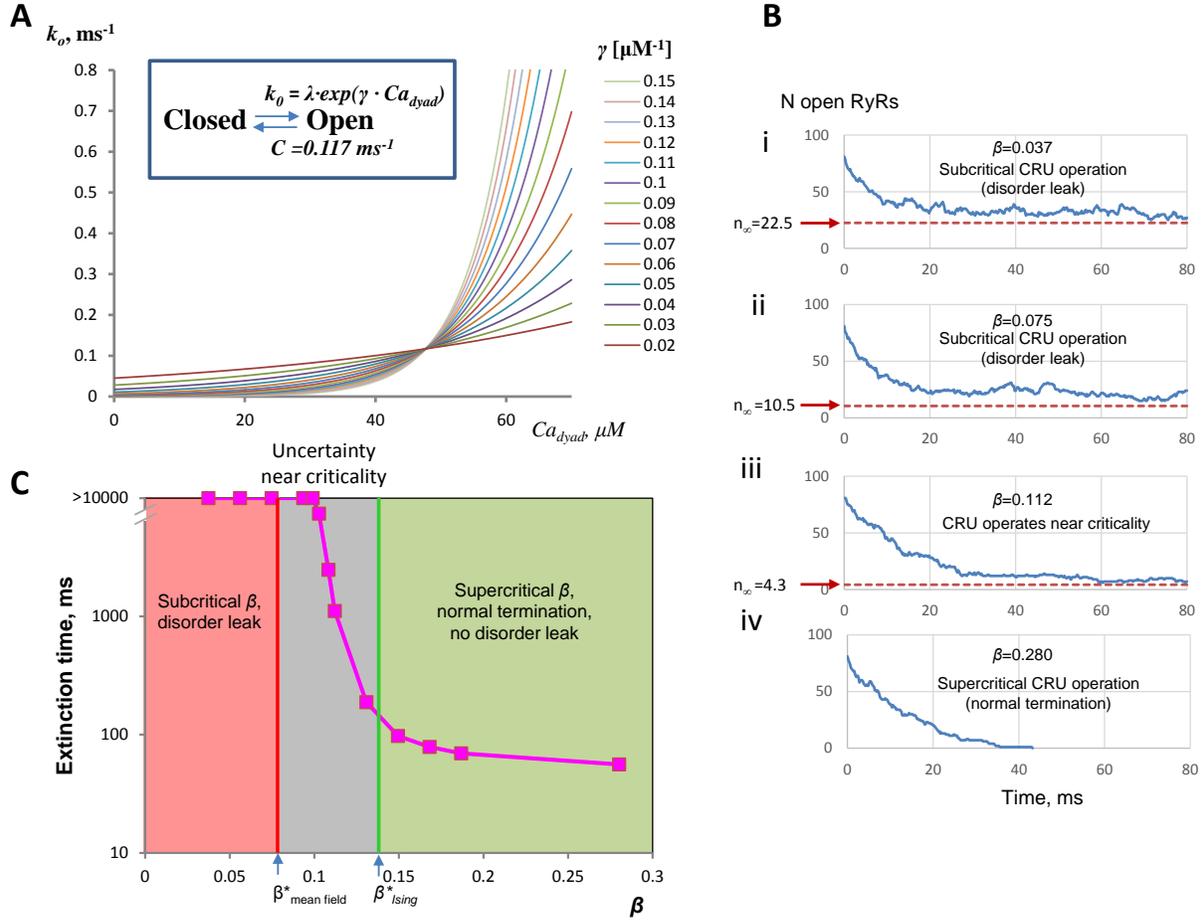

**Fig. 2**: **Onsager's "order-disorder" phase-like transition *via* $\beta$ in a CRU in numerical simulations under $Ca_{JSR}$ clamp**. (A), Plots of opening rate ($k_o$) as a function of $Ca_{dyad}$ with our choices of $\gamma$, $\lambda$, $C$ (Table S3) that keep $h$=0 but allow $\beta$ to vary in a wide range including near criticality. Inset shows our RyR gating scheme. (B), Examples of spark simulations with phase-like transition *via* $\beta$ and emergence of disorder leak in subcritical regime. Dash lines show the low-bound estimates for open number of RyRs at a steady state ($n_\infty$) in supercritical and near-critical regimes. (C), "Order-disorder" phase-like transition in terms of median extinction times (100 sparks for each data point) vs. $\beta$. Arrows show lower and upper bounds estimates for $\beta^*$ obtained analytically ($\beta^*_{mean\_field}$ and $\beta^*_{Ising}$). The phase-like transition happens in the model within these bounds (grey area) that defines near-critical area.

## *Subcritical regime ($\beta<\beta^*$): stochastic attrition and disordered leak*

Our model simulations show a substantial leak in subcritical and near-critical regimes (Fig. 2B) when $h$ is near 0. We further explored the nature of this leak and estimated its level based on RyR parameters. In the subcritical regime RyRs do not effectively interact and their gating becomes disordered (happening effectively independently of each other). The leak then depends almost entirely on the open probability ($P_o$) of RyR at a steady state. If $P_o$ is very small (compared with the inverse of the number of RyRs in a CRU), the spark will terminate *via* pure stochastic attrition; if $P_o$ is relatively high, then the spark will decay to a steady-state level with disordered RyR activity generating the disorder type leak. In general, a lower bound on $P_o$ can be obtained by neglecting interactions and assuming that channels operate independently. In this case, $P_o$ is given as balance between opening and closing rates (Fig. S5A): $P_o= (\lambda/C)/(1 + \lambda/C)$. Please note that $P_o \approx \lambda/C$ at low $\lambda/C$. The probability that all channels ($N_{RyR}$) close, i.e. sparks terminate *via* stochastic attrition is given as $P_{all\_closed}=(1-P_o)^{N_{RyR}}$ (Fig. S5B) and the lower bound estimate of average number of open RyR creating the leak will be $n_\infty = P_o * N_{RyR}$. The respective levels of $n_\infty$ closely describe a lower bound for the number of open RyRs after spark decay in our spark simulations in both subcritical and near-critical regimes (Fig. 2B, panels i, ii, and iii).

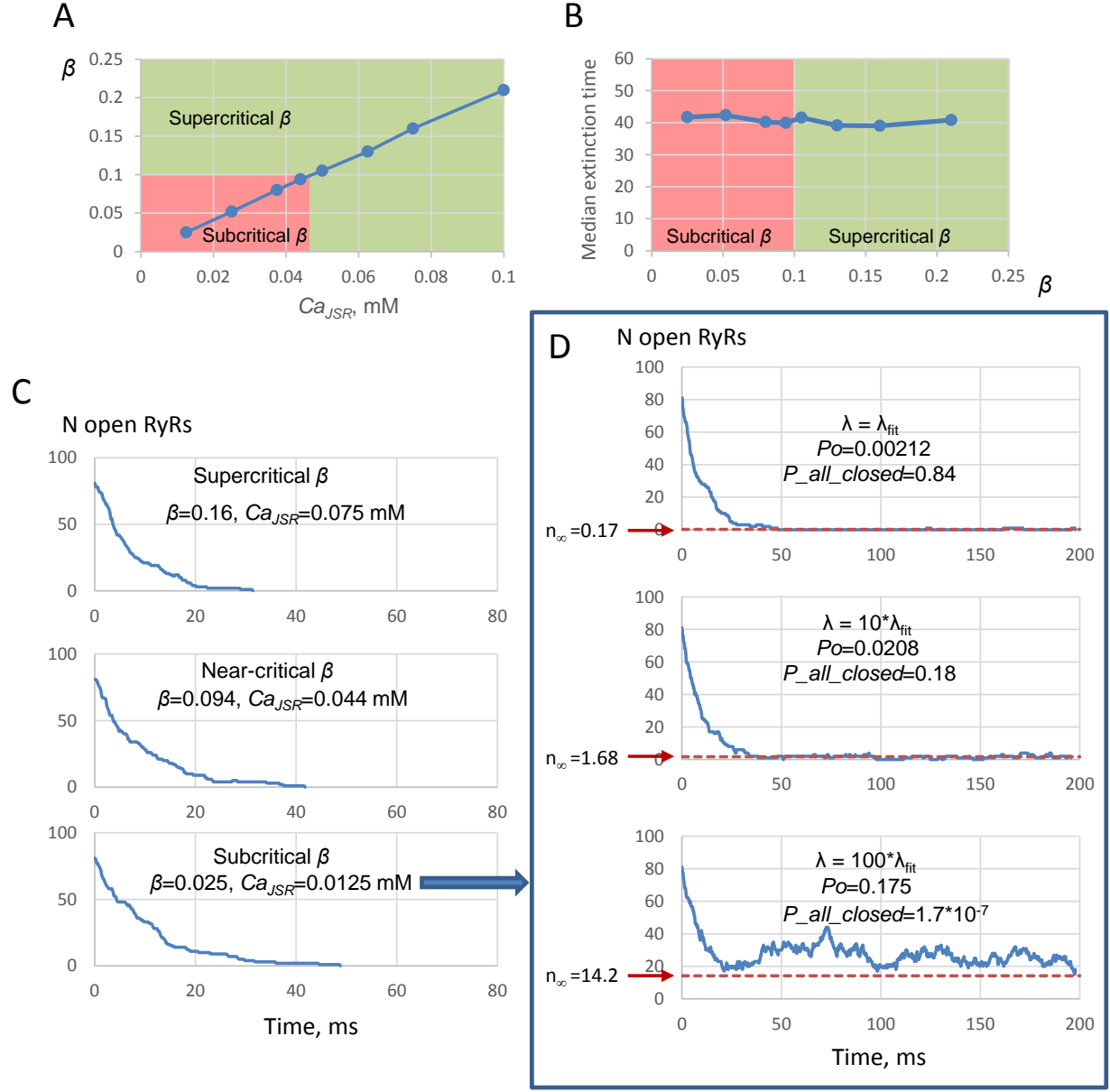

**Fig. 3: Spark termination *via* stochastic attrition in subcritical *β* regime for a CRU with experimentally measured RyR parameters.** (A), Plot shows our choice of $Ca_{JSR}$ levels for SR Ca clamp that define respective values of *β*. Subcritical *β* area is marked by red. The respective values of *h* are given in Table S4. (B), Median extinction times vs. *β* (n=100 sparks for each data point). (C), Examples of spark termination in subcritical, near-critical, and supercritical ranges. (D), Simulations illustrating disorder leak as *λ* increases by 10x and 100x fold from its normal (experimentally measured) value $\lambda_{fit}$. $P_o$, and $P\_all\_closed$ were calculated as shown in Fig. S5. Dash lines show the low-bound estimates for open number of RyRs at a steady state ($n_\infty$).

For RyRs with $\lambda_{fit}$ =0.0002482 ms$^{-1}$ that was fitted to experimental data (Fig. S2C) and with $C$=0.117 ms$^{-1}$ we obtain $P_o$=0.002117. This yields a relatively low $n_\infty$=0.17 and high $P_{all\_closed}$= 0.842 (for 9x9 RyR grid). It means that such sparks even in subcritical regime will likely terminate (*via* stochastic attrition). We tested this prediction by numerical simulations, in which we achieved low $\beta$ range by clamping $Ca_{JSR}$ at relatively low levels, decreasing unitary current ($i_{RyR}$), i.e. the current *via* one RyR. Indeed, the median extinction time was normal (i.e. short, near 40 ms) throughout both subcritical and supercritical regimes of $\beta$ (Fig. 3A-C). On the other hand, RyR generate substantial leak in our simulations in subcritical regime as $\lambda$ increases yielding higher $P_o$ and $n_\infty$ (Fig. 3D).

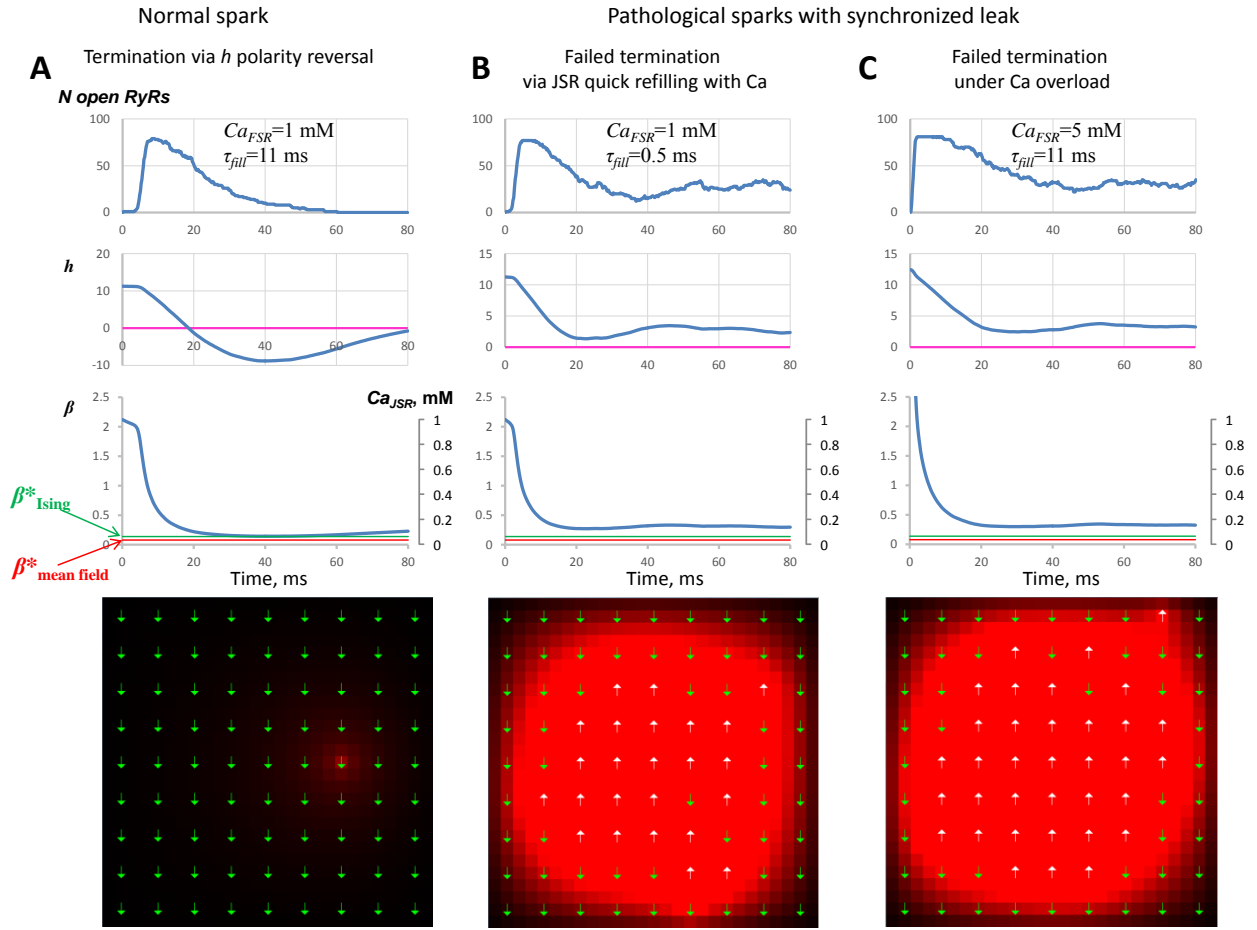

**Fig. 4**: **Examples of numerical simulations for normal and failed spark termination in supercritical $\beta$ regime in a free-running JSR**. (A), A normal spark terminated by phase-like transition *via* h polarity reversal (Movie S1). (B and C), Spark termination fail with JSR quick refilling rate or under Ca overload. Each panel shows dynamics of open RyR number, *h*, *β*, $Ca_{JSR}$. Red and green lines show upper-bound and lower-bound estimates of $\beta^*$ ($\beta^*_{Ising}$=0.138 and $\beta^*_{mean\_field}$=0.0784, respectively). Key parameters ($Ca_{FSR}$ and $\tau_{fill}$) are shown at the traces. Bottom panels show respective distributions of RyR states at *t*=80 ms, i.e. when sparks usually terminate. White and green arrows show open and closed RyRs. $Ca_{dyad}$ is coded by red shades: pure red is 30 μM, black is 0. See also Movies S2 and S3.

*Supercritical regime (β> β\*)*

In the range of *β>β\** the CRU behavior is determined by the sign of *h*. When *h* is positive, RyRs strongly interact *via* CICR and remain mainly open (i.e. generating Ca leak), but when *h* is negative, RyRs tend to close, shown by blue and green areas, respectively in phase diagram in Fig. 1. We have previously demonstrated that the transition from an all-open to an all closed state is sharp and is analogous to the phase transition in ferromagnets on *h* reversal (18). This transition can be illustrated by generating sparks when $Ca_{JSR}$ is clamped at various levels, which yield the respective range of *h* calculated *via* Equation 1. Fig. S6 shows an example of such *h* calculation for 9x9 RyR grid and simulations of the phase-like transitions when *h* reverses its sign.

*Complex realistic CRU behaviors featuring "free-running" SR*

So far, we demonstrated basic steady-state properties of CRU operation using a reductionist approach, i.e. at a steady state when the Ca SR content was clamped. Under physiological conditions the content of JSR ($Ca_{JSR}$) is obviously not fixed: Ca is released *via* RyRs and refilled from FSR *via* a diffusional resistance (parameter $\tau_{fill}$ in the model). The realistic CRU behaviors with "free-running" SR can now be understood and interpreted in terms of the phase diagram in Fig. 1 that summarizes basic CRU properties gleaned from our above reductionist studies. To illustrate the utility of our approach, we simulated emergence of different types of Ca leaks with "free-running" SR. In these simulations *β* was calculated as a (linear) function of $Ca_{JSR}$ according to Equation 2 with *γ* fixed at its physiological value $\gamma_{fit}$.

One simple example is normal spark termination in supercritical regime when JSR Ca content becomes depleted to the level that *h* reverses its sign (18) (green area in phase diagram Fig. 1, see also an example in Fig. 4A and Movie S1). Thus, the leak can occur when *h* simply fails to change its sign. In this case the system does not undergo a phase transition and spark termination fails. In our phase diagram in Fig. 1 this leaky CRU behavior is shown by blue color. This happens when $i_{RyR}$ remains large enough to maintain CICR among neighboring RyRs. The interacting RyRs remain partially synchronized in time and space, which keeps spark alive. Thus, any factor favoring a larger or more sustained $i_{RyR}$ will facilitate the synchronized leak, resulting in long-lasting sparks. For example, one such factor is how quickly JSR is refilled with Ca from free SR (FSR) (parameter $\tau_{fill}$). Another factor is a higher FSR Ca concentration ($Ca_{FSR}$), mimicking Ca overload. Both factors generate leak *via* synchronized RyRs as they both impede SR from depletion and therefore sustain substantial $i_{RyR}$. These leaky CRU behaviors are shown in respective examples of our simulations (in Fig. 4B and C (see also Movies S2, S3) in which we used the RyR opening rate parameters of *λ* and *γ* ($\lambda_{fit}$ and $\gamma_{fit}$) fitted to experimental data obtained by Laver et al. (19) under physiological conditions (Fig. S2C) but different (i.e. abnormal) refilling rate $\tau_{fill}$ or FSR Ca (resulting in Ca overload). Our analytical estimates on *β\** indicate that both normally terminating sparks and this type of synchronized leak occur in our model simulations in supercritical regime, because *β* remains always above the upper-bound *β\** estimate (shown by green lines in Fig. 4).

We also generated different spark behaviors with free-running SR as *λ* increases (Fig. 5). We found different degree of RyR synchronization and respective different behaviors of the system with respect to the criticality at *β\** that can be interpreted using our phase diagram in Fig. 1. Synchronized activity resulted in normal spark termination at supercritical *β* as *h* changed its sign; partially synchronized (self-synchronized) activity resulted in oscillations near criticality (Movie S4); and unsynchronized, completely disordered RyR activity caused leak in subcritical regime (Movie S5).

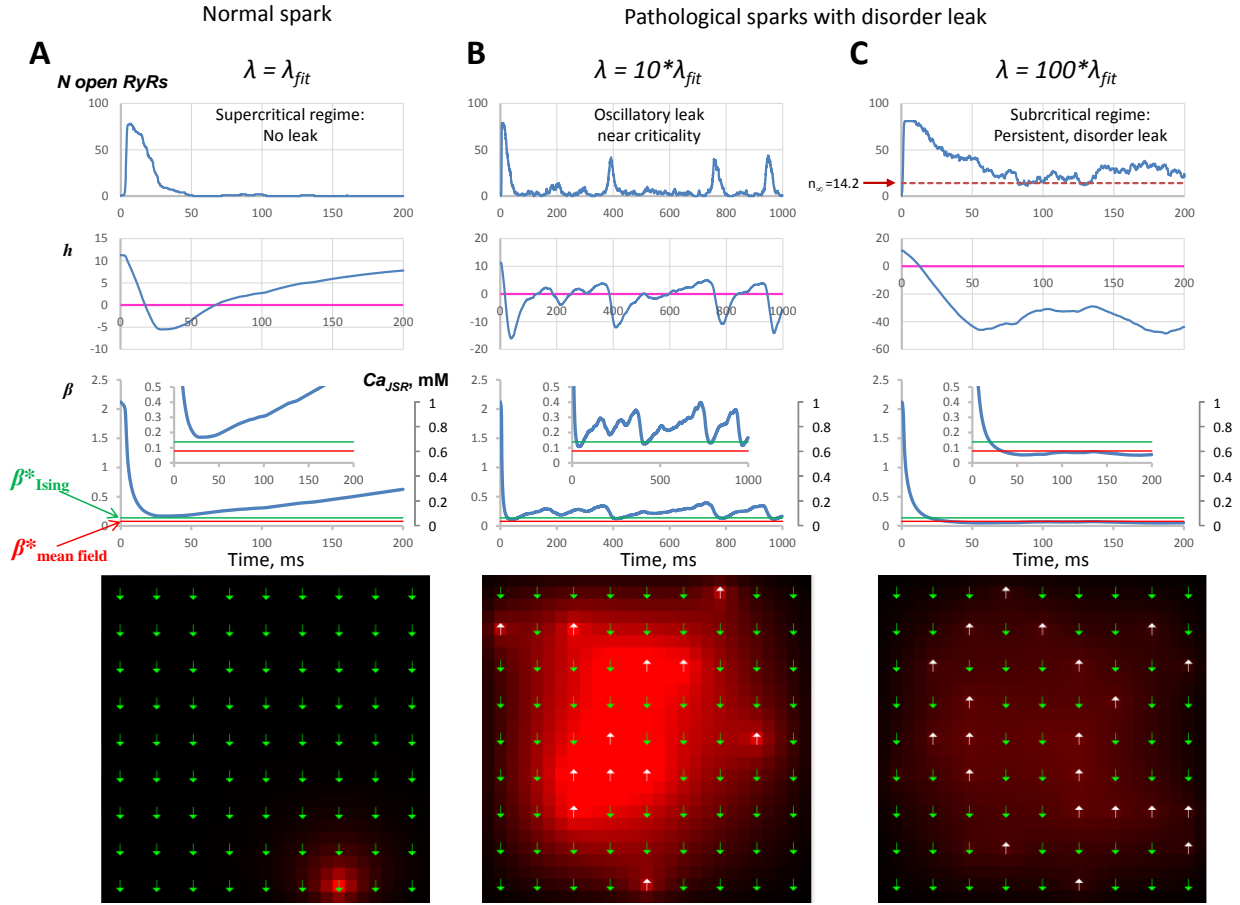

**Fig. 5**: **Examples of pathological sparks with disorder leak emerging at higher $\lambda$ (i.e. higher $P_o$) in a free-running JSR**. (A), Normal supercritical spark with no leak. (B), Oscillatory leak near $\beta$ criticality. (C), Persistent, disorder leak in subcritical regime. Respective $\lambda$ values are shown as multiple of $\lambda_{fit}$ (x1, x10, x100). Red and green lines show upper-bound and lower-bound estimates of $\beta^*$ ($\beta^*_{Ising}$=0.138 and $\beta^*_{mean\_field}$=0.0784, respectively). Dash line in Panel C shows the low-bound estimate for open number of RyRs at a steady state ($n_\infty$). Insets show $\beta$ dynamics with respect to criticality in a more detailed y scale. Bottom panels show respective distributions of RyR states and $Ca_{dyad}$ at $t$=200 ms and $Ca_{dyad}$ (red is 30 μM, black is 0). See also Movies S4 and S5.

## Peierl's contour distinguishes disorder leak and synchronized leak

Here we have described two means of spark termination failure. In one, the magnetic field remains too high and synchronization of open RyRs remains substantial (Fig. 4B,C). In the other, termination fails because of low $\beta$ results in effectively independent RyR openings (Fig. 5C). These two mechanisms can be distinguished by comparing the lengths of Peierls contours (22) of the open RyRs in the grid. Peierls contours are used in statistical physics as a measure of disorder. We constructed the contours in our CRU model as a set of borders between open and closed RyRs (the borders between neighboring open RyRs were excluded). The contour length reflects the degree of disorder in openings among neighboring RyRs, in extreme case representing a checkerboard. In contrast, the smaller contour length indicates a higher local synchronization of opening among neighboring RyRs.

We computed the lengths of Peierls contours in two series of simulations performed for spark failure of different types for the same grid size (see examples in Fig. 6). We obtained different patterns: Open RyRs in supercritical, synchronized leak tend to form big clusters, whereas in disorder leak RyR openings appear disorganized. These different patterns are clearly reflected by Peierls contour length, being about twice longer for disordered leak (Fig. 6B, and Movies S6 and S7).

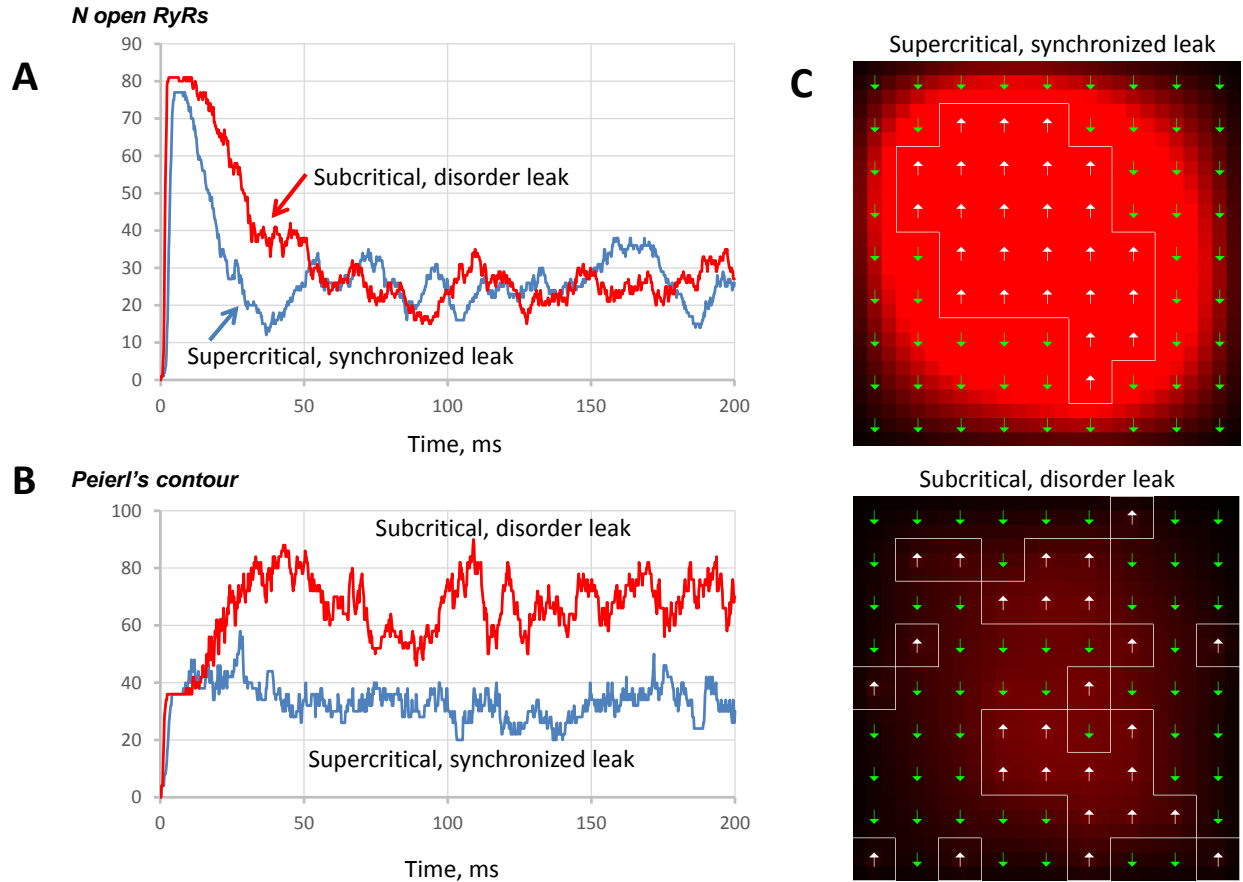

**Fig. 6**: **Peierls contour distinguishes disorder leak and synchronized leak.** (A,B), Time course of number of open RyRs and respective length of Peierls contour in spark termination failure *via* two different mechanisms: blue curve shows Ca leak in supercritical regime (synchronized leak, $\tau_{fill}$=0.5); red curve shows Ca leak in subcritical regime (disorder leak, $\lambda$=100*$\lambda_{fit}$). (C), CRU states at time 200 ms with $Ca_{dyad}$ (red is 30 μM, black is 0). Peierl's contours are shown by white lines over open RyR clusters. See also Movies S6 and S7.

**Discussion**

We approached the problem of abnormal SR Ca leak by application of numerical modeling and statistical mechanics to the CRU viewed as an ensemble of RyRs interacting *via* CICR. Based on our previous finding that the spark termination can be described by a phase transition in an Ising model (Fig. 4A, Movie S1) (18), here we explored and classified the CRU behaviors when spark termination fails and generates the abnormal leak. We summarized our results of numerical and analytical modeling with respect to the *β-h* interplay in Fig. 1. CRU can undergo two types of phase transitions: *via h* polarity reversal and *via β* that is Onsager's "order-disorder" transition (21).

The magnitude of *β*, particularly the binary fact of whether $β < β^*$ or $β > β^*$ determines whether the RyR interactions are important. When $β > β^*$, RyRs states synchronize. In this case, the magnitude of *h*, and again the binary fact of whether *h* is positive or negative, becomes the determining factor for all the RyR states at steady-state. Thus, when *h* changes from positive to negative, spin orientations flip from plus to minus and correspondingly RyRs also flip their states collectively from open to closed (in case of spark termination). In fact, the dividing line between these regimes is identical to the segment of discontinuity of magnetization in the Ising model (see e.g. Fig. 1.2 of Baxter (24)).

On the other hand, in physics, $β < β^*$ corresponds to a high-temperature regime. The high temperature brings disorder by decreasing the interactions of spins to the point that they cease being important. We have demonstrated that a similar phenomenon occurs in the CRU as the analogue of *β* quantity changes, with a phase-like transition as it passes its critical value $β^*$. The RyRs become effectively independent. Once this happens, the magnetic field *h* again plays a key role in RyR behavior. However, when *β* is subcritical, there is no phase transition in *h*. On the contrary, as *h* decreases, the steady-state number of open channels slowly decreases ($β<β^*$, red shaded area). In the CRU model, this disordered regime happens at low SR Ca loads, with $i_{RyR}$ and hence RyRs interactions *via* CICR being reduced. The magnitude of the disorder leak is determined by $P_o$, i.e. the leak increases at high opening rates and/or low closing rates (as a ratio of $λ/C$ increases, Fig. S5). At very low $P_o$ sparks still terminate in the subcritical regime *via* pure stochastic attrition (lower almost white area in Fig. 1). This type of spark termination does not seem to be physiological as it happens at extremely low SR loads (Fig. 3C, bottom panel).

In supercritical regime ($β>β^*$) the leak can occur when *h* fails to change its sign (remaining within blue area in Fig. 1). In this case partially synchronized RyR openings persist resulting in long-lasting sparks. This synchronized leak happens when the SR does not sufficiently deplete and RyRs keep interacting *via* CICR. More efficient connectivity of JSR with FSR or larger SR Ca loads facilitates the leak (Fig. 4B,C and Movies S2 and S3). Conditions for both synchronized leak and disorder leak are now defined deterministically by simple analytical formulas (dashed lines in Fig. 1 for *h*=0 and $β^*$).

*Different leak types: distinctive patterns and different functional consequences*

While both leak regimes may look similar in terms of average number of open RyRs, they have a different nature, causing different system behavior. In simulations they appear differently to the eye as locally synchronized openings vs. noisy/disordered RyR firing. The two regimes can be also distinguished objectively and quantitatively by calculating Peierls contour lengths (22) (Fig. 6, Movies S6 and S7). The Peierls contour length is solely an output parameter and can provide deep insight into the mechanism of leak even in cases when the input parameters are unavailable.

We found different levels of synchronization of RyR openings: synchronized for long-lasting sparks (Fig. 4), unsynchronized for disorder leak and partially synchronized (self-synchronized) oscillatory activity near criticality (Fig. 5). Long-lasting sparks with highly synchronized RyR activity extend towards diastole and can therefore contribute to initiation of arrhythmia *via* EADs or DADs. Ca overload and high connectivity of JSR with FSR support persistent high-amplitude $i_{RyR}$ and thus facilitate synchronized leak (*via* long-lasting sparks) that may provoke Ca waves and DADs (3) and sometimes also EADs (25). A long-lasting spark is more likely to trigger a Ca wave than a normal short-lived spark simply because of the longer exposure of the neighboring CRU to the Ca released by the longer spark. Similarly, once the wave is initiated, long-lasting sparks on its way will better support the wave propagation. This supporting role of long-lasting sparks in wave generation and propagation has been discussed earlier by Brochet et al. (9).

The partially synchronized oscillatory activity of RyRs near criticality (Fig. 5B, Movie S4) is reminiscent of Ca embers (26). Such embers and spontaneous oscillatory sparks can also provoke DAD (even EADs if oscillation period is short). On the other hand, this type of spontaneous oscillatory releases during diastole may be involved in normal pacemaker function *via* so-called "calcium clock" (27-30).

Disorder leak is expected to deteriorate contraction amplitude of heart muscle due to low SR Ca levels associated with this leak type (Fig. 5C). However, the cause-effect relation between SR Ca level and disorder leak could be tricky. A low SR Ca level is required (but not sufficient) for the disorder leak. For the leak to become notable, $P_o$ should be high. On the other hand, if the disorder leak becomes substantial, it can further deplete SR Ca levels. For example, in our simulations in Fig. 5, $\lambda$ facilitates RyR open state, resulting in substantial Ca depletion of JSR that, in turn, shifts the CRU operation towards criticality and further to disorder leak regime. On the other hand, once CRU operates within disorder regime, the increasing leak *via* $\lambda/C$ is not a phase transition per se. We estimated a lower bound on the expected number of open channels given by the binomial approximation, i.e. larger $\lambda/C$ → larger $P_o$ → more leak (Fig. 3D and Fig. S5).

*Insights into normal CRU operation: RyR interactions must be balanced*

Under physiological conditions $Ca_{JSR}$ becomes depleted during spark termination. As $Ca_{JSR}$ decline, $\beta$ and $h$ change simultaneously with time. With values of $\lambda_{fit}$ and $\gamma_{fit}$ fitted to experimental data from Laver et al. (19) obtained under physiological conditions (Fig. S2C) our simulations show that $\beta$ remains supercritical throughout the duration of the spark, i.e. it remains above our upper estimate of 0.138 (Fig. 4A). This means that CICR interactions between RyRs remain the key aspect of the system operation; and abrupt and robust termination occurs indeed *via* "polarity reversal" phase transition. When $h$ does not reverse, termination fails, and the CRU generates synchronized leak (Fig. 4B,C). Our results also demonstrate that the normal spark termination requires balanced RyR interactions. Since in our model $h$ and $\beta$ are inversely related, high RyR interactions yield both a high supercritical $\beta$ and a positive $h$ through the duration of the spark. This combination does not allow termination as both interactions of open RyRs remain strong and the RyR preferred state (indicated by the sign of $h$) is to be open. However, RyR interactions that are too low (e.g. *via* low SR Ca load) would yield a low $\beta$ which makes termination less robust (white area in Fig. 1), i.e. prone to substantial disorder leak at higher $P_o$ or near-critical oscillations with uncertain outcome (Fig. 5B).

*Different leak types require different treatments*

The two leak types will require different treatments to shift leaky CRU operation towards normal spark termination. One possibility for effective treatment of synchronized leak would decrease the unitary current $i_{RyR}$, e.g. by decreasing Ca overload. On the other hand, treatments of the disorder leak would target the number of open RyRs or RyR gating to decrease $P_o$. This approach would at least help to convert disorder leak into spark termination *via* stochastic attrition (faint red to white in Fig. 1). But to fully normalize spark termination *via* phase transition (green area in Fig. 1), we have to increase $\beta$ (i.e. RyR interactions). This is the opposite to synchronized leak treatment. Moreover, treatments must be delicate and balanced to get exactly into the normal spark termination, avoiding overtreatment shifts from one pathological leak to the other (red and blue areas).

Another important insight offered by our study is that the transitions from one regime to another (and respective treatments) are abrupt, *via* the respective criticalities shown by dash lines in Fig. 1. It means that (i) a treatment may remain ineffective if its effect remains within the same functional paradigm (same color in Fig. 1), (ii) it can be uncertain if we reach the near-critical regimes including spontaneous oscillations (Fig. 5B), (iii) the effect can be abrupt when we cross the criticality and shift into another functional paradigm (another color). Thus, our leak classification and exact mathematical formulas for the boundaries of different functional paradigms can be helpful in developing and optimizing treatments of different types of leak to restore normal CRU operation and normal heart function.

*Our results are in line with prior studies*

Prior numerical simulation studies characterized the invisible releases (sub-sparks and quarks) and sparks that fail to terminate (11-17). The invisible releases are similar to disorder leak in subcritical $\beta$ regime reported here in our model. The prior reports that long-lasting sparks are facilitated by rapid inter- and intra-SR Ca diffusion (15,16) or faster JSR refilling rate with Ca (13) is in line with our results (Fig. 4B). In general, we show here that any factors increasing and supporting unitary current $i_{RyR}$ would prevent phase transition *via h* polarity reversal and facilitate a synchronized leak (the system remains locked in blue area in Fig. 1).

Because Ca sparks are rare events during diastole, smaller "'nonspark'' events (often called invisible releases) are thought to be responsible for a major part of SR Ca leak (31). The occurrence of spark-mediated leak and non-spark (invisible) leak was experimentally studied by Zima et al. (32). They showed that spark-mediated leak sharply increases as the SR Ca load surpasses 500 μM, whereas non-spark leak dominates at lower loads. We interpret the experimentally measured non-spark leak as a manifestation of disorder leak (in our terms) that occurs in our model at low SR Ca loads, when CICR is negligible and RyRs cannot effectively interact. At higher loads, CICR generates sparks, whereas the non-spark leak becomes saturated (32) (see their Fig. 3A). This saturation can be explained by increased amplitude of $i_{RyR}$ (i.e. single RyR current) at higher SR loads that is expected to compensate the smaller total number of non-spark RyR openings (Nopen), so that the product of the two ($i_{RyR}$*Nopen) defining the net Ca flux would tend to sustain. An additional factor that supports non-spark leak at higher loads could be also oscillating tails of "quarky" releases (see next section).

*Ca Quarks and Quarky Ca releases*

The term of Ca quark was introduced by Lipp and Niggli in 1996 (33) when they claimed to resolve experimentally openings of individual RyR within a CRU triggered by flash photolysis of caged Ca. The term "quarky calcium release" as a new type of Ca leak contributing to "invisible" leak was introduced in 2011 by Brochet et al. (9). Their measurements resolved a finer structure of sparks and showed that some sparks consist of an initial stereotypical high-flux release followed by a highly variable "quarky" release that is attributable to CICR. By this description, this type of CICR-driven leak is similar to oscillatory leak in a CRU near criticality (in our terms), see Fig. 5B and Movie S4. Brochet et al. speculated that quarky release occurs because of CRU imperfections including rogue RyRs or subclusters of RyRs (Fig. 7 in their paper (9)). On the other hand, the present study shows that a quarky release can also occur within a perfect RyR grid.

*Limitations and Future directions*

Our model features a perfect lattice of RyRs (Fig. S1). In reality the CRU geometry features various sizes and morphologies (34) and some (e.g. phosphorylated) RyRs may have different gating properties. Furthermore, we cover here leaks emerging within a CRU. Other leak types, e.g. spontaneous sparks and waves, may have different mechanisms.

Our numerical model of spark (13) is reduced to a minimal gating scheme of one open and one closed state (Fig. 2A, inset), i.e. RyRs interact exclusively *via* CICR. Other possible interactions, such as allosteric coupling, are not considered in our reductionist approach here. Furthermore, the real Ca near couplon during cell contraction will vary, depending on numerous factors, such as Ca fluxes *via* RyRs in neighboring CRUs, L-type Ca channels, Na/Ca exchanger, Ca pumps, cytosol Ca buffers, Ca diffusion in cytosol, etc. Thus, while cell Ca signaling/dynamics is clearly a multiscale problem, our reductionist study focuses on a single spark model at a background cytosolic Ca level near 100 nM.

Future directions:

1. To the best of our knowledge, our approach offers a new view and terminology on the problem of Ca leak in the heart. In addition to the mathematical clarity of the model in simplified reductionist settings, our formulas provide an effective computational tool to aid in detailed numerical modeling of Ca dynamics and high-dimensional computationally intensive parametric search. Knowledge of just two aggregate parameters $h$ and $\beta$ in conjunction with our phase diagram yields hitherto unknown instantaneous information about the system's steady-state (i.e. where the system "wants" to be).

2. The Ising model and other systems at and near criticality exhibit very special behaviors, quite different from their behaviors away from criticality. First off, the time of convergence to equilibrium is long. This means in terms of RyRs that if we start with all RyRs open, the termination time might be long even if $h$ is negative, making criticality particularly dangerous. Secondly, the spin correlations exhibit long range order, i.e. event sizes are large and exhibit heavy tails. Lastly, event sizes (closely related to spin-to-spin correlation length) near criticality in every known magnetic material obey a universal formula (with "universal exponents"), and we would expect it to also hold in the CRU. All these near-critical phenomena merit further exploration.

3. Our mapping to the Ising model gives us a formula for the steady-state probability of every configuration of open/closed RyRs. While this formula is not computationally tractable, the ratio of probabilities of any two possible configurations is easy to obtain. This yields powerful information that has not truly been utilized and remains to be understood.

4. The CRU oscillatory behaviors (e.g. in Fig. 5B) are similar to spontaneous Ca oscillations found in pacemaker cells, i.e. "calcium clock" (27-30) In terms of statistical physics, this may be analogous to superheating, which can be explored using hysteresis in the Ising model.

5. Recent results (34) indicate that RyR positioning is not a perfect grid. There are three main differences that may impact system behavior: the small deviations of most RyRs away from the perfect grid positions, missing RyRs, and the non-trivial non-square shape of the whole cluster. Each of these differences merits further study, with different methodology. The small deviations could be modeled as disorder in the interaction function. The missing RyRs would correspond to a dilute Ising model (35) which is currently actively studied in mathematical physics. The interesting shapes of clusters require a careful analysis of how system behavior depends on the shape of its boundary.

Lastly, the Ising model is a powerful and universal tool recently introduced into this field. We hope it will spur a new way of thinking about the clinical problem of Ca leak in mathematical terms. Thus, a key envisaged future direction would be to classify possible interventions in terms of their effects on Ising parameters $h$ and $\beta$ and to use this knowledge in designing new effective therapies. Our finding that leak emerges abruptly as underlying parameters change only a little (as shown in Fig. 1) will have important implications for drug design and dosage.

*Conclusions*

The present study considered the mechanisms of the pathological SR Ca leak as an emerging, macroscopic phenomenon, i.e. generated by a statistical ensemble of interacting individual RyRs within a CRU. Using Ising model from statistical mechanics we discovered two types of phase transitions and two types of emergent leak. One leak type is a synchronous release by RyRs which strongly interact *via* CICR (e.g. in Ca overload). Another type is a disorder leak with no effective RyR interactions (e.g. at low SR levels). The disorder leak is distinguished from synchronized leak by larger Peierls contour lengths. Normal spark termination and leak regimes are summarized in a *β-h* phase diagram (Fig. 1), which can guide future treatments to inhibit leak and normalize Ca release by a CRU.

Our numerical simulations and Ising model separately and altogether demonstrate the two major points: 1) Any factor that increases $i_{RyR}$ (unitary current *via* one RyR) will increase RyR interactions *via* CICR and prevent spark termination, facilitating Ca leak. These factors include JSR quick refilling with Ca and Ca overload (Fig. 4B and Fig. 4C, respectively). 2) When $i_{RyR}$ is small and channels do not interact, the leak is determined simply by opening rate of a single channel (i.e. Po in terms of ion channel gating) (Fig. 3D and Fig. 5C).

**Author Contributions**
A. V. Maltsev: designed research; performed research; wrote the manuscript; contributed analytic tools; analyzed data.
M. D. Stern: designed research; contributed numerical model tools.
V. A. Maltsev: designed research; performed research; wrote the manuscript; analyzed data; developed visualization software.

**Acknowledgments**
The work was supported by the Intramural Research Program of the National Institute on Aging, National Institute of Health. AM acknowledges the support of the Royal Society University Research Fellowship UF160569.

**Supporting Citations**
References (36-38) appear in the Supporting Material.

**References**
1. Fabiato, A. 1983. Calcium-induced release of calcium from the cardiac sarcoplasmic reticulum. Am J Physiol 245:C1-14.
2. Bers, D.M. 2014. Cardiac sarcoplasmic reticulum calcium leak: basis and roles in cardiac dysfunction. Annu Rev Physiol 76:107-127.
3. Boyden, P.A., and G.L. Smith. 2018. $Ca^{2+}$ leak-What is it? Why should we care? Can it be managed? Heart Rhythm 15:607-614.
4. Niggli, E., N.D. Ullrich, D. Gutierrez, S. Kyrychenko, E. Polakova, and N. Shirokova. 2013. Posttranslational modifications of cardiac ryanodine receptors: $Ca^{2+}$ signaling and EC-coupling. Biochim Biophys Acta 1833:866-875.
5. Eisner, D.A., J.L. Caldwell, K. Kistamas, and A.W. Trafford. 2017. Calcium and Excitation-Contraction Coupling in the Heart. Circ Res 121:181-195.
6. Franzini-Armstrong, C., F. Protasi, and V. Ramesh. 1999. Shape, size, and distribution of $Ca^{2+}$ release units and couplons in skeletal and cardiac muscles. Biophys J 77:1528-1539.
7. Stern, M.D. 1992. Theory of excitation-contraction coupling in cardiac muscle. Biophys J 63:497-517.
8. Cheng, H., W.J. Lederer, and M.B. Cannell. 1993. Calcium sparks: elementary events underlying excitation-contraction coupling in heart muscle. Science 262:740-744.
9. Brochet, D.X., W. Xie, D. Yang, H. Cheng, and W.J. Lederer. 2011. Quarky calcium release in the heart. Circ Res 108:210-218.
10. Cheng, H., and W.J. Lederer. 2008. Calcium sparks. Physiol Rev 88:1491-1545.
11. Williams, G.S., A.C. Chikando, H.T. Tuan, E.A. Sobie, W.J. Lederer, and M.S. Jafri. 2011. Dynamics of calcium sparks and calcium leak in the heart. Biophys J 101:1287-1296.
12. Sato, D., and D.M. Bers. 2011. How does stochastic ryanodine receptor-mediated Ca leak fail to initiate a Ca spark? Biophys J 101:2370-2379.
13. Stern, M.D., E. Rios, and V.A. Maltsev. 2013. Life and death of a cardiac calcium spark. J Gen Physiol 142:257-274.
14. Walker, M.A., G.S. Williams, T. Kohl, S.E. Lehnart, M.S. Jafri, J.L. Greenstein, W.J. Lederer, and R.L. Winslow. 2014. Superresolution modeling of calcium release in the heart. Biophys J 107:3018-3029.
15. Sato, D., T.R. Shannon, and D.M. Bers. 2016. Sarcoplasmic Reticulum Structure and Functional Properties that Promote Long-Lasting Calcium Sparks. Biophys J 110:382-390.
16. Song, Z., A. Karma, J.N. Weiss, and Z. Qu. 2016. Long-Lasting Sparks: Multi-Metastability and Release Competition in the Calcium Release Unit Network. PLoS Comput Biol 12:e1004671.
17. Wescott, A.P., M.S. Jafri, W.J. Lederer, and G.S. Williams. 2016. Ryanodine receptor sensitivity governs the stability and synchrony of local calcium release during cardiac excitation-contraction coupling. J Mol Cell Cardiol 92:82-92.
18. Maltsev, A.V., V.A. Maltsev, and M.D. Stern. 2017. Clusters of calcium release channels harness the Ising phase transition to confine their elementary intracellular signals. Proc Natl Acad Sci U S A 114:7525–7530.
19. Laver, D.R., C.H. Kong, M.S. Imtiaz, and M.B. Cannell. 2013. Termination of calcium-induced calcium release by induction decay: an emergent property of stochastic channel gating and molecular scale architecture. J Mol Cell Cardiol 54:98-100.
20. Gillespie, D., and M. Fill. 2013. Pernicious attrition and inter-RyR2 CICR current control in cardiac muscle. J Mol Cell Cardiol 58:53-58.

21. Onsager, L. 1944. Crystal Statistics. I. A Two-Dimensional Model with an Order-Disorder Transition. Physical Review 65:117-149.
22. Peierls, R. 1936. On Ising's model of ferromagnetism. Proceedings of the Cambridge Philosophical Society 32:477-481.
23. Aizenman, M., and R. Fernandez. 1988. Critical Exponents for Long-Range Interactions. Letters in Mathematical Physics 16:39-49.
24. Baxter, R.J. 1989. Exactly Solved Models in Statistical Mechanics. Academic Press.
25. Huffaker, R., S.T. Lamp, J.N. Weiss, and B. Kogan. 2004. Intracellular calcium cycling, early afterdepolarizations, and reentry in simulated long QT syndrome. Heart Rhythm 1:441-448.
26. Csernoch, L. 2007. Sparks and embers of skeletal muscle: the exciting events of contractile activation. Pflugers Arch 454:869-878.
27. Lakatta, E.G., V.A. Maltsev, and T.M. Vinogradova. 2010. A coupled SYSTEM of intracellular $Ca^{2+}$ clocks and surface membrane voltage clocks controls the timekeeping mechanism of the heart's pacemaker. Circ Res 106:659-673.
28. Maltsev, A.V., V.A. Maltsev, M. Mikheev, L.A. Maltseva, S.G. Sirenko, E.G. Lakatta, and M.D. Stern. 2011. Synchronization of stochastic $Ca^{2+}$ release units creates a rhythmic $Ca^{2+}$ clock in cardiac pacemaker cells. Biophys J 100:271-283.
29. Stern, M.D., L.A. Maltseva, M. Juhaszova, S.J. Sollott, E.G. Lakatta, and V.A. Maltsev. 2014. Hierarchical clustering of ryanodine receptors enables emergence of a calcium clock in sinoatrial node cells. J Gen Physiol 143:577-604.
30. Maltsev, A.V., Y. Yaniv, M.D. Stern, E.G. Lakatta, and V.A. Maltsev. 2013. RyR-NCX-SERCA local crosstalk ensures pacemaker cell function at rest and during the fight-or-flight reflex. Circ Res 113:e94-e100.
31. Santiago, D.J., J.W. Curran, D.M. Bers, W.J. Lederer, M.D. Stern, E. Rios, and T.R. Shannon. 2010. Ca sparks do not explain all ryanodine receptor-mediated SR Ca leak in mouse ventricular myocytes. Biophys J 98:2111-2120.
32. Zima, A.V., E. Bovo, D.M. Bers, and L.A. Blatter. 2010. $Ca^{2+}$ spark-dependent and -independent sarcoplasmic reticulum $Ca^{2+}$ leak in normal and failing rabbit ventricular myocytes. J Physiol 588:4743-4757.
33. Lipp, P., and E. Niggli. 1996. Submicroscopic calcium signals as fundamental events of excitation--contraction coupling in guinea-pig cardiac myocytes. J Physiol 492 ( Pt 1):31-38.
34. Jayasinghe, I., A.H. Clowsley, R. Lin, T. Lutz, C. Harrison, E. Green, D. Baddeley, L. Di Michele, and C. Soeller. 2018. True Molecular Scale Visualization of Variable Clustering Properties of Ryanodine Receptors. Cell Rep 22:557-567.
35. Bodineau, T., B. Graham, and M. Wouts. 2013. Metastability in the dilute Ising model. Probability Theory and Related Fields 157:955-1009.
36. Levin, D.A., Y. Peres, and E.L. Wilmer. 2008. Markov Chains and Mixing Times. American Mathematical Society.
37. Martinelli, F. 1999. Lectures on Glauber Dynamics for Discrete Spin Models. *In* Lectures on Probability Theory and Statistics. Bernard P, editor. Springer-Verlag, Berlin Heidelberg.
38. Evans, M. Lecture Notes. Section 11: Exact results for the Ising Model https://www2.ph.ed.ac.uk/~mevans/sp/sp10.pdf.

**Biophysical Journal**

**Supplemental Information**

# Mechanisms of Calcium Leak from Cardiac Sarcoplasmic Reticulum Revealed by Statistical Mechanics

Anna V. Maltsev[1], Michael D. Stern[2], Victor A. Maltsev[2]*

[1]School of Mathematics, Queen Mary University of London,
London, United Kingdom

[2]Laboratory of Cardiovascular Science, Biomedical Research Center, Intramural Research Program, National Institute on Aging, NIH, Baltimore, Maryland, USA.

# 1. ISING MODEL METHODS (modified from (1))

**1.1. Brief Introduction to the Ising Model.** The Ising model we will work with consists of binary random variables (i.e. taking values ±1) called **spins** positioned on a 2D finite grid $\Lambda$ (e.g. section 3.3.5 in (2)). A configuration of spins is a function $\sigma$ that assigns 1 or -1 to each point $x \in \Lambda$. The configuration space $\Omega$ is the set of all possible assignments of spins to points in $\Lambda$, i.e. all possible functions $\sigma : \Lambda \to \{1, -1\}$. A **interaction profile** $\phi: \mathbb{R} \to \mathbb{R}$ is a function with $\phi(x) \to 0$ rapidly as $x \to \infty$ and $\phi > 0$. We choose $\phi$ so that $\phi(1) = 1$. We furthermore place our finite grid $\Lambda$ inside of a bigger grid $\Lambda_b$ (b for boundary) and let $\sigma(x) = -1$ for any $x \in \Lambda_b \setminus \Lambda$. In this way we impose a -1 **boundary condition** on $\Lambda$. Here $\Lambda_b \setminus \Lambda$ must "frame" $\Lambda$ and its thickness has to be at least as wide as the effective interaction range, which in our case will be around 5. To be precise, if $\Lambda$ is a *n* by *m* grid, $\Lambda_b$ will be a *n+10* by *m+10* grid with $\Lambda$ situated in the middle of $\Lambda_b$. The Hamiltonian is

$$[1] \qquad H(\sigma) = -\sum_{x,y \in \Lambda_b} \phi(|x-y|)\sigma(x)\sigma(y) - h\sum_{x \in \Lambda_b} \sigma(x)$$

Here the first sum is over $\Lambda_b$ instead of $\Lambda$. This is necessary to ensure the interaction with the boundary.

In physics, *h* is the magnetic field. The Hamiltonian can be interpreted as the energy of the system. The equilibrium measure (Gibbs measure) is given by

$$[2] \qquad \pi(\sigma) = Z^{-1} e^{-\beta H(\sigma)} .$$

The normalization constant *Z* is well-defined since our lattice $\Lambda$ is finite, and we will not need to know it explicitly for our analysis. Here $\beta$ is the inverse temperature. (For further information on the general Ising model, of which this is an instance, cf Sections 2.1 and 2.2 in (3)).

**1.2. Dynamic Ising: Detailed Balance and the Transition Rates.** Let $\Lambda$ be a 2 dimensional integer grid of a finite size. Recall that $\Omega$ is the configuration space and let $\sigma: \Lambda \to \{1, -1\}$ be an element of $\Omega$. One can introduce a dynamic on spin configurations so that the configuration space $\Omega$ becomes the state space for a Markov chain with a transition matrix $P$. We introduce the notation $\sigma^x$ to mean

$$\sigma^x = \begin{cases} \sigma(y) & \text{for } y \neq x \\ -\sigma(y) & \text{for } y = x \end{cases}$$

i.e. $\sigma^x$ coincides with $\sigma$ everywhere except at $x$, where the spin is reversed. To obtain a Glauber-like dynamic for the Ising model, it suffices to choose a spin uniformly at random at each time increment and to give the probability that it flips, i.e. to give $P(\sigma \to \sigma^x)$. y

The condition on $P$ that guarantees that $\pi$ as in [2] is indeed the equilibrium measure for the Markov chain is called detailed balance, and it states that the Markov chain is reversible with respect to $\pi$ (cf equation (1.30) and Proposition 1.19 in (2)). The equation for detailed balance is the following: for all $\sigma \in \Omega$ and $x \in \Lambda$ we have that

[3]
$$P(\sigma \to \sigma^x)e^{-\beta H(\sigma)} = P(\sigma^x \to \sigma)e^{-\beta H(\sigma^x)}$$

This is equivalent to

[4]
$$\frac{P(\sigma \to \sigma^x)}{P(\sigma^x \to \sigma)} = e^{\beta H(\sigma) - \beta H(\sigma^x)}$$

$$= e^{-2\beta(\sum_{y \in \Lambda_b} \phi(|x-y|)\sigma(x)\sigma(y) + h\sigma(x))} = e^{-2\beta\sigma(x)(\sum_{y \in \Lambda_b} \phi(|x-y|)\sigma(y) + h)}$$

The detailed balance equations will be satisfied for a wide variety of rates $P$, so we can choose $P$ to be most appropriate to our CRU model. Since we know that the release channel opening rate is an exponential while the closing rate is a constant, we look for $P$ so that the transition from -1 to 1 is exponential while the transition from 1 to -1 is a constant. This indeed can be achieved simultaneously with the detailed balance condition. If $\sigma(x) = -1$ we let

$$P(\sigma \to \sigma^x) = Ce^{2\beta\left(\sum_{y \in \Lambda_b} \phi(|x-y|)\sigma(y)+h\right)}$$

yielding that $P(\sigma^x \to \sigma) = C$ to satisfy detailed balance. Thus, the Markov chain is given as follows. We pick a location $x$ uniformly at random, and define the transition matrix $P$ to be:

[5] $$P(\sigma, \sigma^x) = \begin{cases} Ce^{2\beta(\sum_{y \in \Lambda_b} \phi(|x-y|)\sigma(y)+h)} & \text{for } \sigma(x) = -1 \\ C & \text{for } \sigma(x) = 1 \end{cases}$$

Here time is continuous and the above are transition rates. In our numerical model, time is discrete and we take $\Delta t = 0.05$ ms. The transition matrix with the discretized time becomes

[6] $$P(\sigma, \sigma^x) = \begin{cases} \Delta t C e^{2\beta(\sum_{y \in \Lambda_b} \phi(|x-y|)\sigma(y)+h)} & \text{for } \sigma(x) = -1 \\ \Delta t C & \text{for } \sigma(x) = 1 \end{cases}$$

and we ensure that $\Delta t$ is small enough so that all transition probabilities are smaller than 1. Letting also $P(\sigma, \sigma) = 1 - P(\sigma, \sigma^x)$ ensures that $P$ is indeed stochastic.

**1.3. The CRU as an Ising Model.** A numerical model of the CRU consists of a square grid of Ca release channels $\Lambda$ and each release channel can be open or closed. We assign 1 to each open and -1 to each closed release channel, thus obtaining a configuration $\sigma : \Lambda \to \{1, -1\}$. We introduce the constant $U$ to represent the spatial distance between nearest release channels. In our numerical model, is $U = 30$ nm.

We let $\psi$ be the 1D slice of the time-stable spatial Ca profile resulting from the opening of one release channel. This is sufficient to contain all the information about the Ca profile since $\psi$ is rotationally symmetric. We obtain $\psi$ from our numerical simulation. However, $\psi$ is an immediate result of the environment, including current, diffusion, and buffer and is not an emergent property. We interpret it as a scaled interaction profile, and let $\phi$ in [1] be given as $\phi(r) = \psi(Ur)/\psi(U)$, where $Ur$ is the distance to the open release channel. The multiplication by $U$ accounts for the fact that the release channels are $U$ units apart while spins are 1 unit apart. The division by $\psi(U)$ is a choice of scaling for the interaction profile function $\phi$. With this scaling we have $\phi(1) = 1$. We choose this scaling for $\phi$ so that at the nearest neighbors its value matches the classic Ising model, where each spin interacts with 4 neighbors with a strength of 1.

The distance between CRUs is assumed to be too large for Ca from one CRU to influence another. On the other hand, Ca is diffusing out of the CRU and in this way the release channels in the CRU interact with the outside. The model would be identical if the CRU were surrounded by release channels that are always closed. In this way, the boundary condition of the CRU model is equivalent to a negative boundary condition of the Ising model.

We will compute the analogues of inverse temperature $\beta$ and the magnetic field $h$ in our CRU model as functions of initial model parameters. They play the exact same role in the mathematical description of our CRU model as they do in the Ising model even though they do not carry the same physical meaning. We will note that $\beta$ is an increasing function of the concentration of Ca inside the junctional SR and we vary the SR Ca in our numerical model to test the predictions of the CRU Ising model.

**1.4. Relating [Ca] and the Ising Hamiltonian.** Let us introduce the set $S(x) := \{s \in \mathbb{R} : s = |x - y| \neq 0 \text{ for some } y \in \mathbb{Z}^2\}$. We can rewrite both the local [Ca] at $x$ (we denote it $[Ca](x)$) and the exponent in the -1 to 1 transition in $P$ in terms of a sum over $S(x)$. Given a configuration of open and closed release channels $\sigma$ and a given release channel at a point $x$, let $N_{Us}$ be the number of open RyRs at a distance $Us$ from $x$. If the release channel at $x$ is closed, we can approximate [Ca] at $x$ by

$$[Ca](x) = \sum_{s \in S(x)} \psi(Us) N_{Us} \qquad [7]$$

We similarly rewrite P. We introduce the following notation: $T_s(x) :=$ total number of spins at distance s from $x$; $L_s(x) :=$ number of -1 spins at distance s from $x$; $N_s(x) :=$ number of +1 spins at distance s from $x$; and we have $N_s(x) + L_s(x) = T_s(x)$.

Henceforth in this section, let us fix a site $x \in \Lambda$ and suppress the dependence on $x$ in $T_s$, $L_s$, $N_s$, and $S$ for ease of notation. Then we can rewrite the expression in the exponent of the Ising -1 to +1 transition probability in [5] in the following way:

$$[8] \qquad \sum_{y \in \Lambda_b} \phi(|x-y|)\sigma(y) = \sum_{s \in S} \phi(s)(N_s - L_s) = \sum_{s \in S} \phi(s)(2N_s - T_s)$$
$$= 2\sum_{s \in S} \phi(s) N_s - \sum_{s \in S} \phi(s) T_s \approx 2\sum_{s \in S} \phi(s) N_s - 2\pi \int_{s>.5} \phi(s)\, ds$$

In the last approximate equality, we have replaced $\sum_{s \in S} \phi(s) T_s$ by $2\pi \int_{s>.5} \phi(s) ds$ where the factor of $2\pi$ is due to the fact that $\sum_{s \in S} \phi(s) T_s$ is approximately a 2D integral of a rotationally symmetric function. We observe that the first term in the final expression in [8] is a scalar multiple of the total Ca [Ca] $(x)$ as given in [7].

**1.5. Crucial Parameters and the Spark Termination Criterion.** We want to solve for the analogues of $h$ and $\beta$ in the CRU model. We again fix a site $x \in \Lambda$ and suppress the dependence on $x$ in [Ca] and $S$ for ease of notation. From experimental data we fit the exponential $\lambda e^{\gamma[Ca]}$ to the Ising transition rate from -1 to +1 in [5]:

$$\lambda e^{\gamma[Ca]} = C e^{2\beta(\sum_{y \in \Lambda_b} \phi(|x-y|)\sigma(y) + h)}$$

Then we replace the LHS using [7] and the RHS using the expression derived in [8] to obtain

[9a]
$$\lambda e^{\gamma \sum_{s \in S} \psi(Us) N_{Us}} = Ce^{2\beta(2\sum_{s \in S} \phi(s)N_s - 2\pi \int_{s>.5} \phi(s)ds + h)}$$
$$= Ce^{-4\beta\pi \int_{s>.5} \phi(s) + 2\beta h} e^{4\beta(\sum_{s \in S} \phi(s)N_s)}$$

Since we wish the above equality to hold for any configuration, we must equate the coefficients of $\sum_{s \in S} \phi(s)N_s$ to obtain

[9b]
$$\beta = \gamma\psi(U)/4.$$

Next we equate the coefficients in front of $e^{4\beta(\sum_{s \in S} \phi(s)N_s)}$ to obtain

$$\lambda = Ce^{-4\beta\pi \int_{r>.5} \phi(s)dr + 2\beta h}$$

yielding that

[10]
$$h = \frac{1}{2\beta}\ln\left(\frac{\lambda}{C}\right) + 2\pi \int_{r>.5} \phi(r)dr$$

Rewriting $h$ in terms of the Ca profile $\psi$ we obtain

[11]
$$h = \frac{2}{\gamma\psi(U)}\ln\left(\frac{\lambda}{C}\right) + 2\pi \int_{r>U/2} \frac{\psi(r)}{U\psi(U)}dr$$

Since $h$ is the analogue of the magnetic field in the CRU model, the emergent behavior of release channels can be predicted based on $h$. During termination all the release channels begin in an open state (analogous to +1). The Ca diffusion out of CRU is equivalent to a negative boundary condition. We can hence deduce the **signal termination criterion**: If $h < 0$, then the spark will terminate and this termination is mathematically identical to reversal of polarity in ferromagnetism. Mathematically, this phase transition follows from the Lee-Yang theorem. On the other hand, if $h > 0$, the spark will not terminate.

## 2. ESTIMATES OF CRITICAL $\beta$

### 2.1. Lower bound estimate using mean field approach

This calculation closely follows the lecture notes of Prof. Martin Evans from University of Edinburgh (4).

The Ising configuration energy is

$$E(\sigma) = -h \sum_{x \in \Lambda} \sigma(x) - \sum_{x,y \in \Lambda} \phi(|x-y|)\sigma(x)\sigma(y) \qquad (1)$$

Let $\epsilon(\sigma(x))$ be the contributions involving spin $x$ to this energy:

$$\epsilon(\sigma(x)) = -h\sigma(x) - \sigma(x)\sum_{y \in \Lambda} \phi(|x-y|)\sigma(y) = -\sigma(x)(h + \sum_{y \in \Lambda} \phi(|x-y|)\sigma(y))$$

We then replace the contributions from $\sigma(y)$ by their mean values:

$$\epsilon(\sigma(x)) = -\sigma(x)(h + \sum_{y \in \Lambda} \phi(|x-y|)\langle\sigma(y)\rangle) = -h_{mf}\sigma(x)$$

where

$$h_{mf} = h + 2\pi m \int_{s>0.5} \phi(s)ds \qquad (2)$$

and $m = \langle\sigma(y)\rangle$. Then we replace the energy in (1) by the energy of non-interacting spins each experiencing a field with magnitude $h_{mf}$. In this approximation we know the single-spin Boltzmann distribution:

$$P(\sigma(x)) = \frac{e^{-\beta\epsilon_{mf}(\sigma(x))}}{\sum_{\sigma(x)=\pm 1} e^{-\beta\epsilon_{mf}(\sigma(x))}} = \frac{e^{\beta h_{mf}\sigma(x)}}{e^{\beta h_{mf}} + e^{-\beta h_{mf}}} \qquad (3)$$

We now must ensure that the approximation is self-consistent. The mean value of magnetization predicted by (3) should match the mean value used in (2). We obtain the equation:

$$m = \sum_{\sigma(x)=\pm 1} P(\sigma(x))\sigma(x) = \frac{e^{\beta h_{mf}} - e^{-\beta h_{mf}}}{e^{\beta h_{mf}} + e^{-\beta h_{mf}}} = \tanh(\beta h_{mf}) \qquad (4)$$

yielding the mean field equation for magnetization

$$m = \tanh(\beta h + 2\beta\pi m \int_{s>0.5} \phi(s)ds)$$

Setting $h = 0$, we note that for low $\beta$ the equation

$$m = \tanh(2\beta\pi m \int_{s>0.5} \phi(s)ds)$$

has only one solution and for high beta it has 3 solutions. The transition happens when

$$\frac{d}{dm}\tanh(2\beta\pi m \int_{s>0.5} \phi(s)ds)|_{m=0} \geq 1$$

Using a Taylor expansion of tanh near 0, we obtain that $2\beta\pi \int_{s>0.5} \phi(s)ds \geq 1$ yielding that the critical $\beta$ is

$$\beta^* = \frac{1}{2\pi \int_{s>0.5} \phi(s)ds} \approx 0.0784$$

## 2.2. Upper bound estimate using classical 4-neighbor Ising beta critical

We can obtain an upper bound on $\beta^*$ by comparing with the classical nearest neighbor Ising model with

$$J = \frac{\int_{s>0.5} \phi(s)ds}{4} = \frac{12.7586}{4} = 3.19$$

i.e. the total quantity of interactions in our model divided between for nearest neighbors. In the classical Ising model

$$\frac{J}{kT_c} = 0.441$$

for these formulae cf for example (6.2.2) and (6.2.16) of Baxter (5). Since $\beta^* = 1/(kT_c)$, dividing the (5) by J = 3.19, we obtain an upper bound for beta critical of 0.138.

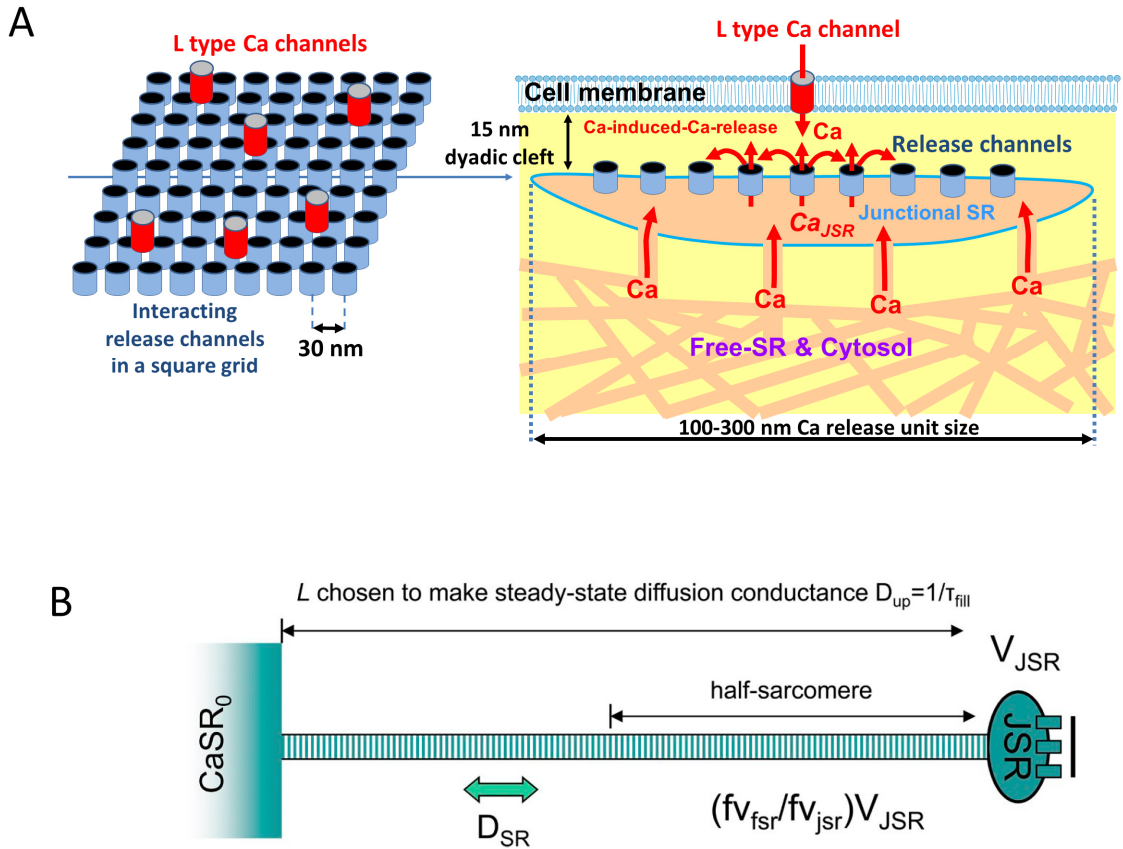

**Fig. S1. A schematic representation of CRU function in cardiac cells in Stern numerical model that describes collective behavior of RyR ensemble in a CRU during spark activation and termination.**
**A**: Illustration of geometry and Ca fluxes in the model. Modified from Maltsev et al. (1). Each RyR operates in 2 states: open and closed, with no time-dependent inactivation, coupled gating, or allosteric interactions. A lattice of 9x9 RyRs separated from each other by U=30 nm is embedded on a JSR that features calsequestrin and a diffusive connection with a free SR (FSR) that is equipped with a Ca pump. A 15 nm dyadic space features Ca buffers and a diffusive connection to the cytoplasm. The model simulates intradyadic local Ca dynamics on a nanoscale, with a voxel size =10x10x15 nm (xyz). Individual RyRs release Ca and interact via CICR. The dyadic space includes physiological Ca buffers and the released Ca diffuses to JSR border to reach the cytoplasm that has a fixed [Ca] of 100 nM. **B**: Diffusional connection between junctional SR (JSR) to free SR (FSR) determining JSR refilling with Ca. The connection is made through a tube of local FSR, whose length and diameter are chosen to match the observed steady-state diffusion resistance (characterized by time constant $\tau_{fill}$) and the observed volume fraction of FSR. For the standard parameters and a true half-sarcomere length of 1 μm, the effective SR tube length is 1.995 μm. From Stern et al. (6). In the numerical simulations, the boundary condition for Ca at the edge of the couplon was determined by adding, to the background cytosolic Ca (100 nM), the product of the flux of Ca leaving the couplon and an estimated diffusion resistance between the boundary and "infinity" in the cytosol. The diffusion resistance estimate was originally determined from analytical computations in cylindrical coordinates assuming a central source and numerical integration of the diffusion equation in rectangular coordinates using PDEase (Macsyma Corp, Arlington, MA).

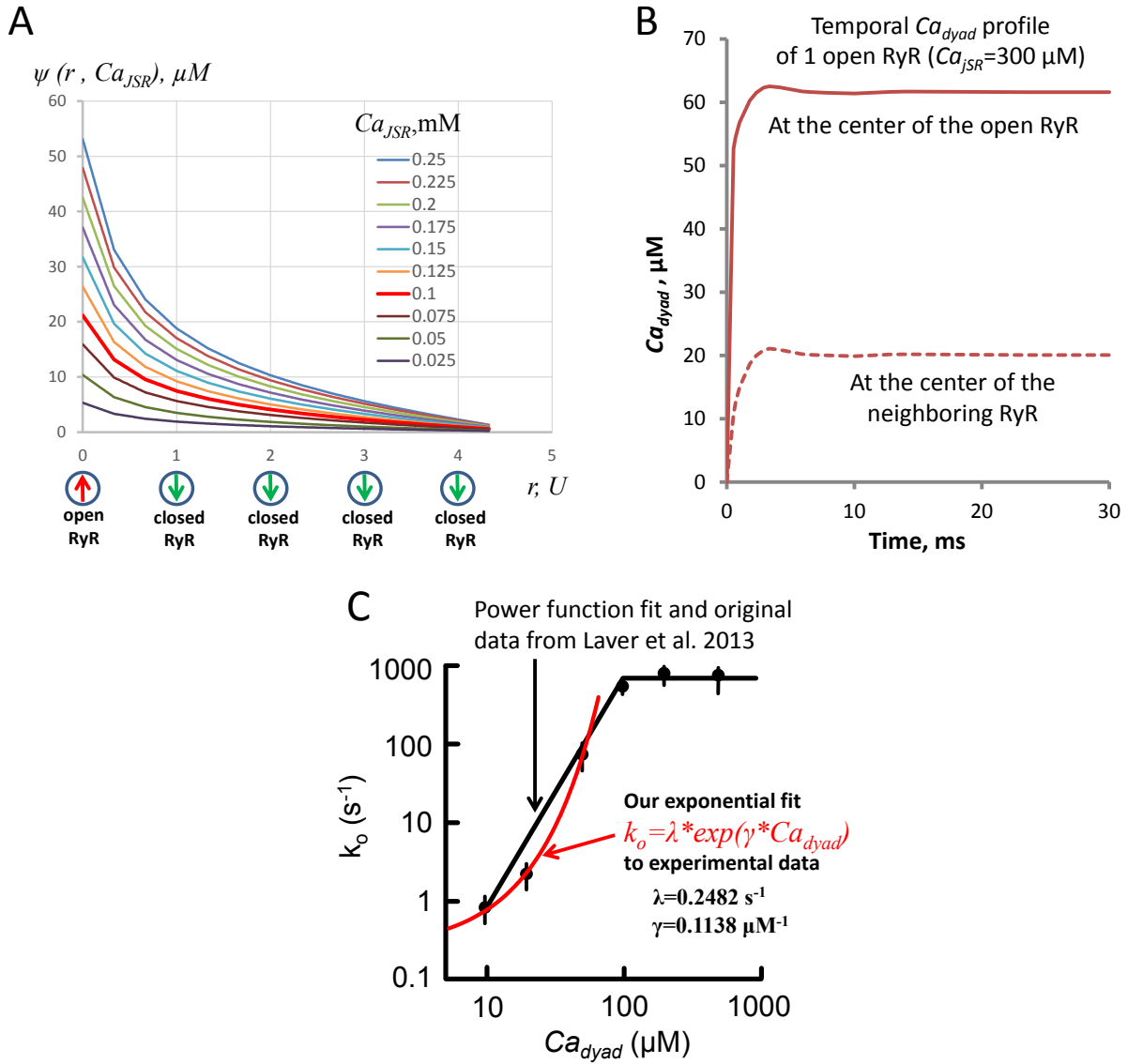

**Fig. S2. Construction of an exact mapping between a CRU described by Stern model and the Ising model of interacting spins.**
**A**, A steady-state spatial $Ca_{dyad}$ profile at various $Ca_{JSR}$ when one RyR is open in the center of 9x9 grid at $r=0$. **B**, Representative $Ca_{dyad}(t)$ when one RyR is open in the center of the grid: at the open RyR and its closest neighbor. **C**, The exponential relation of RyR opening rate vs. $Ca_{dyad}$. All previous models fit a power function to original data obtained in lipid bilayers. Here we fit an exponential (red line) to the same data points (original data and power fit are reproduced from Laver et al. (7). Thus, we replaced the quadratic opening rate in original Stern model with the exponential opening rate from this fit. Modified from Maltsev et al. (1).

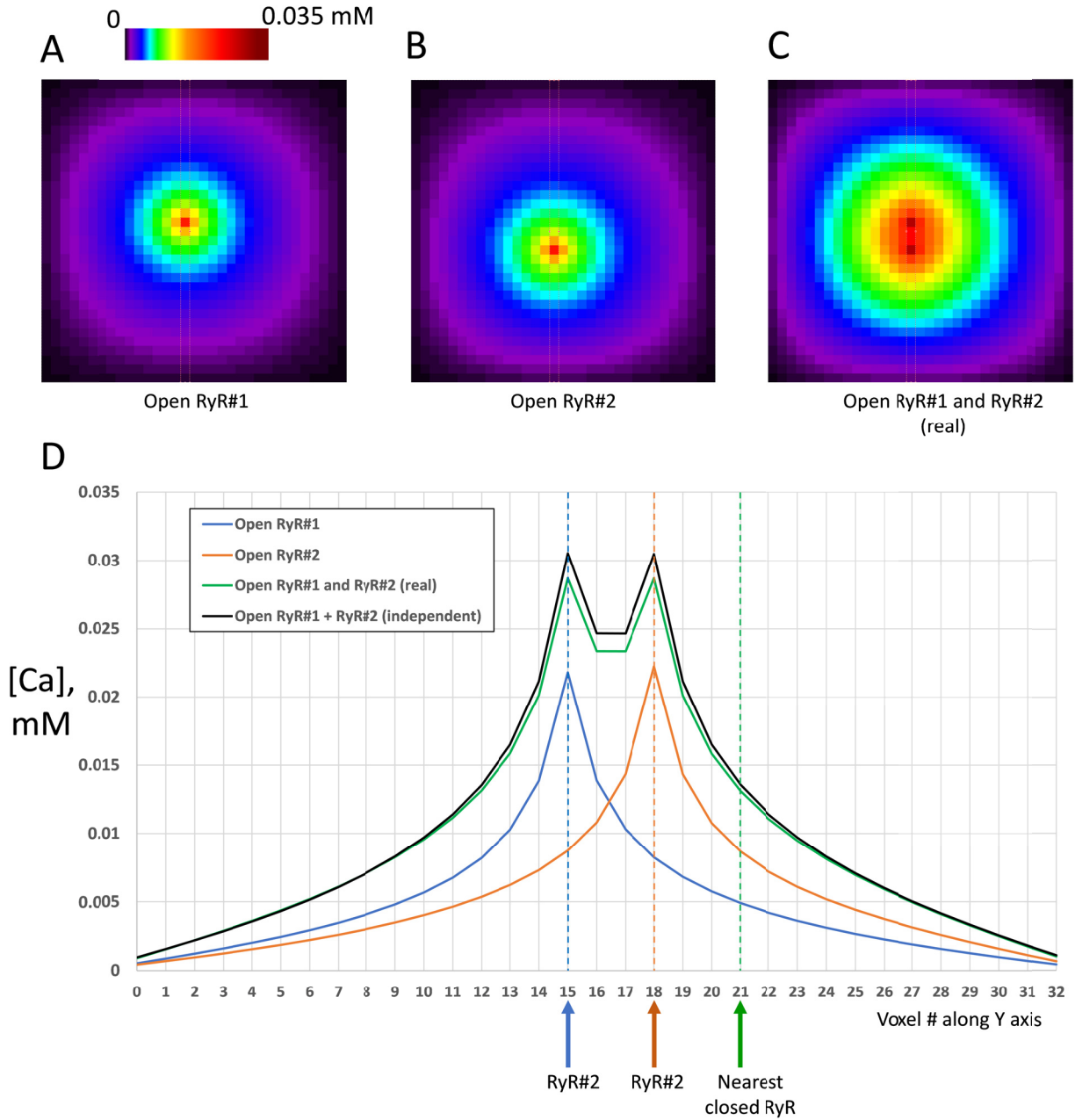

**Fig. S3: Independence of Ca release flux and minor effect of driving force reduction due to open neighboring RyRs.**
**A-C**: Simulations of steady state local Ca distributions in the dyadic space with 2 neighboring channels (RyR#1 and RyR#2) opening separately and simultaneously, respectively, at a fixed Ca SR level of 0.1 mM. **D**: Local Ca profiles through open channels along vertical (Y) axis, respectively. The local profile "RyR#1 + RyR#2 (independent)" was calculated as a formal sum of profiles "Open RyR#1" and "Open RyR#2" as if channels operate independently. The real profile "Open RyR#1 and RyR#2 (real)" generated by numerical simulation deviates negligibly from that formal sum of individual profiles. The deviation at either open channel #1 or #2 position is 5.9% and at the nearest closed neighbor is only 3.5% (arrows). See text and Table S2 for details.

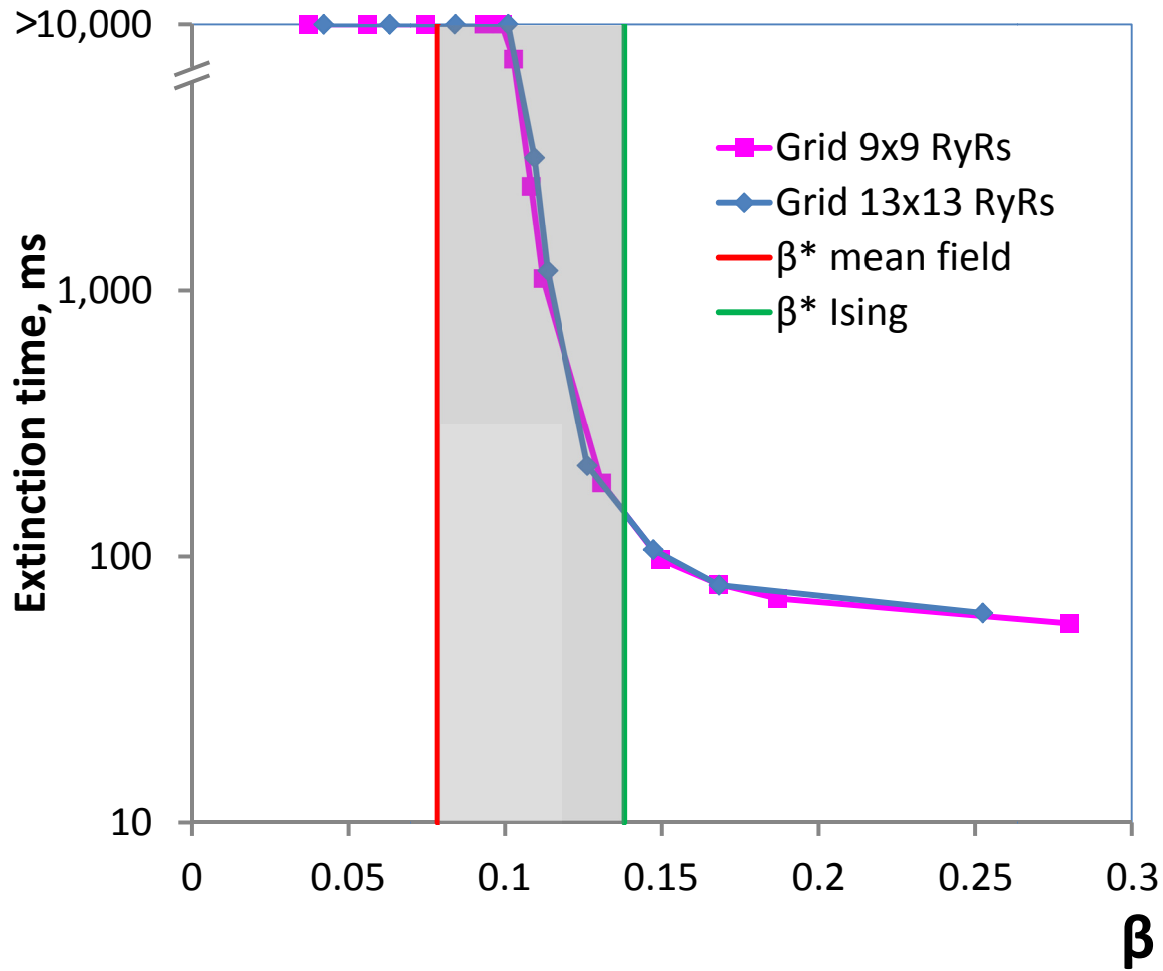

**Fig. S4: "Order-disorder" phase transition occurs within the same β range for a wide range of RyR cluster sizes.**
Shown are "order-disorder" phase transitions for two cluster sizes of 81 and 169 in terms of median extinction times (100 sparks for each data point) vs. β. Vertical lines show lower and upper bounds estimates for β* obtained analytically: β*$_{mean\_field}$ (red line) and β*$_{Ising}$ (green line). The extinction times and respective phase transitions follow basically the same dependence on β within these bounds (grey area) for both cluster sizes. Note: Simulation data for 9x9 cluster size (in magenta) is the same is in Figure 2C of the main text. The magenta and blue curves overlap within the accuracy of our method evaluating 100 sparks for each data point. If we assume that phase transition happens when extinction time becomes >10,000 ms then β* = 0.09899 and 0.10097, for 81 and 169 clusters, respectively, i.e. the difference of 0.00198 is about 2%.

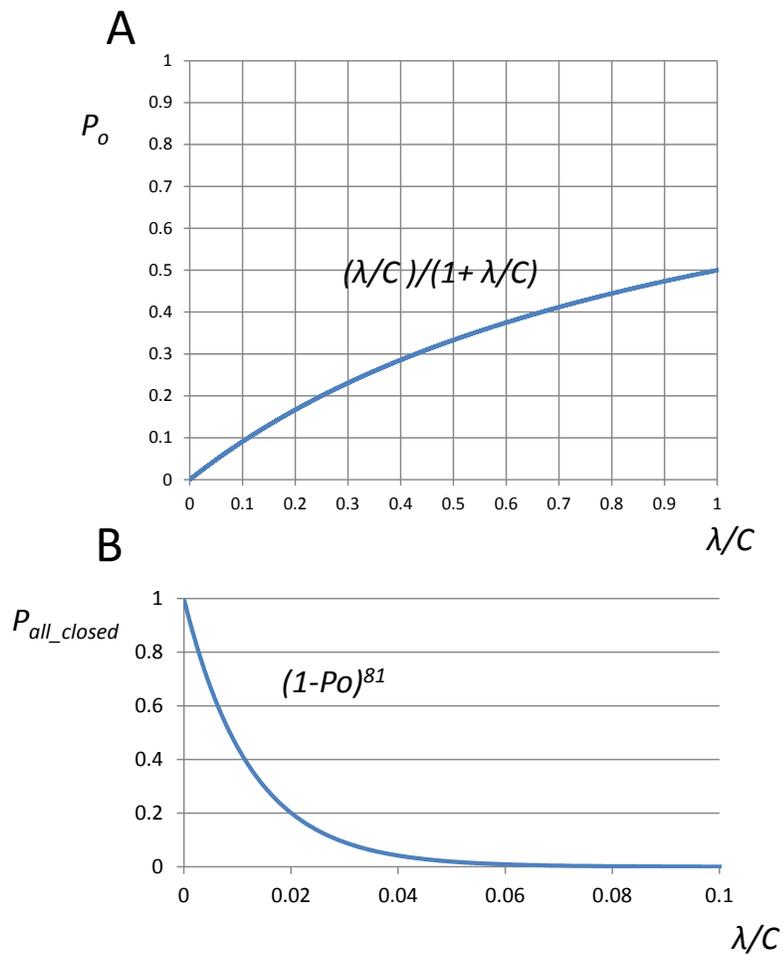

**Fig. S5: Leaky operation of RyRs in subcritical regime neglecting any remaining interactions with each other.**
**A**: Steady-state open probability ($P_o$) is given as balance between opening ($\lambda$) and closing ($C$) rates as $P_o = (\lambda/C)/(1 + \lambda/C)$. **B**: The probability that all 81 RyRs become closed.

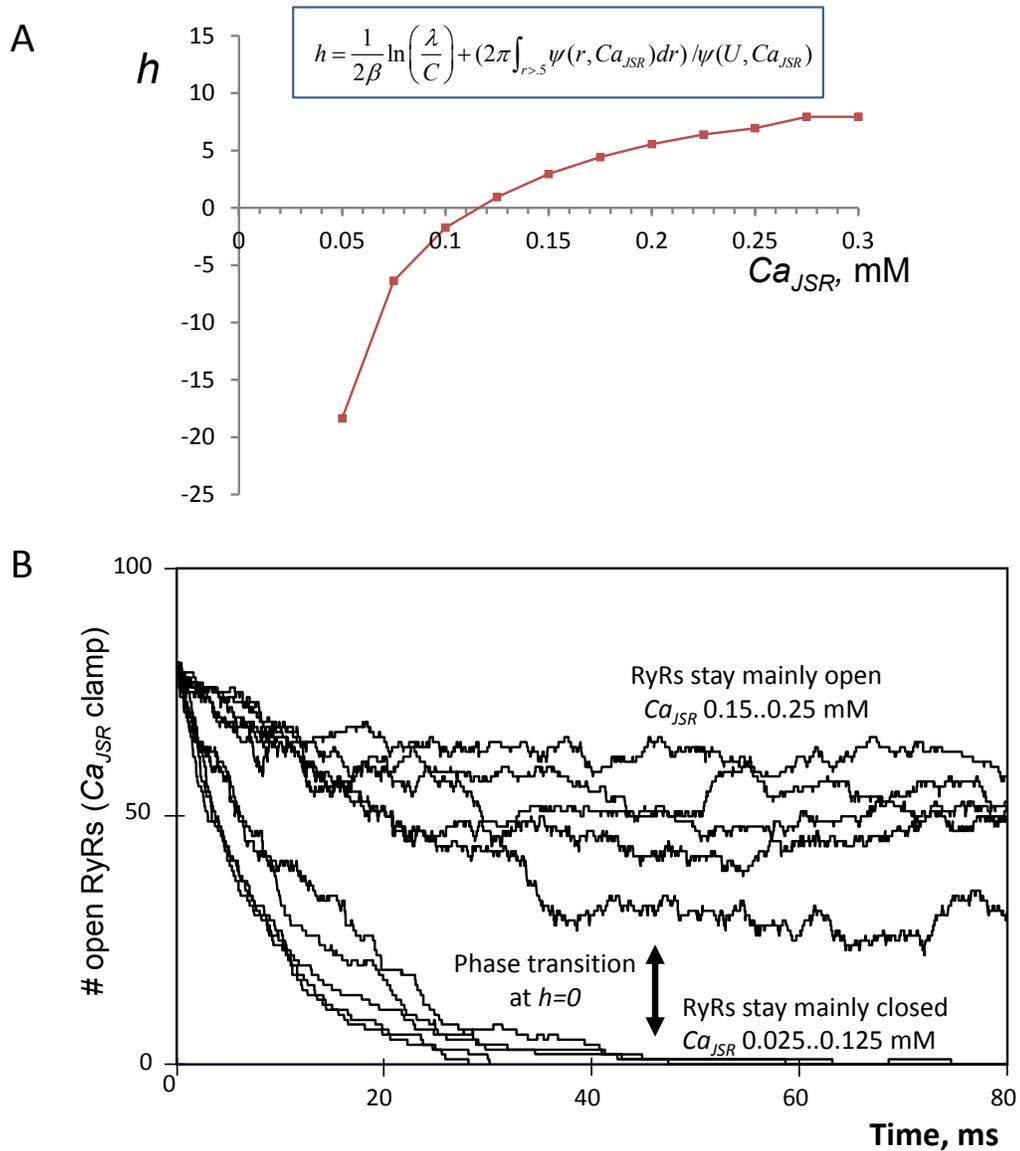

**Fig. S6: Supercritical $\beta$ regime: Phase transition in a CRU of 9x9 RyRs via $h$ reversal as $Ca_{JSR}$ is clamped at various levels in Stern model.**
A: Calculation of $h$ as a function of $Ca_{JSR}$, using modified Equation 11 in supplement (indicated in the inset). In this equation, $\psi(U, Ca_{JSR})$ is interaction profile $\psi(r, Ca_{JSR})$ taken for $r = U = 30$ nm shown by vertical arrow in Online Figure IIA. In turn, $\beta$ was calculated as $\beta=\gamma\psi(U, Ca_{JSR})$. B: Evolution of RyR ensemble at various $Ca_{JSR}$ levels after all RyRs are set in the open state at time 0. RyRs stay mainly closed at $Ca_{JSR}$ below 0.12 mM, but become mainly open above 0.12 mM. The sharp transition in the numerical model behavior is in line with the Ising model prediction of the phase transition at 0.12 mM on $h$ reversal. Modified from Maltsev et al.(1).

**Table S1 – Relationship between the abstract Ising model and ion channels and CRU parameters**

| Abstract Ising model parameters | Equations | Release channel (RyR) parameters | CRU parameters |
|---|---|---|---|
| Inverse temperature $\beta$ | Equation [9b] in the supplement $$\beta = \gamma\psi(U)/4$$ | $\gamma$ is opening rate exponent in $k_o = \lambda * exp(\gamma * Ca_{dyad})$ (Fig. S2C)  $\lambda$ is opening rate scaling factor, i.e. a virtual rate at $Ca_{dyad} = 0$  The single channel conductance*, in Stern spark model UIC=0.35pA, i.e. $i_{RyR}$ at resting $Ca_{JSR,rest} = 1$ mM | $U$=30 nm is the distance between neighboring RyRs  $\psi(r, Ca_{JSR}, UIC)$ is the interaction profile†‡, i.e. $Ca_{dyad}$ at distance $r$ from open channel with a given $Ca_{JSR}$ and single channel conductance (i.e. UIC in the Stern model). |
| Magnetic field $h$ | Equation [11] in the supplement $$h = \frac{2}{\gamma\psi(U)}\ln\left(\frac{\lambda}{C}\right) + 2\pi\int_{r>U/2}\frac{\psi(r)}{U\psi(U)}dr$$ | | |

Notes:
* The single channel conductance is present in the equations implicitly via $\psi$
† $\psi$ also depends on the CRU size (it tends to decrease for very small CRUs due to boundary effects).
‡ The present study does not provide an analytical formula for $\psi$; rather it was a read-off from numerical model simulations (Fig. S2A).

**Table S2 – Independence of Ca release flux via neighboring RyRs, i.e. minor effect of driving force reduction due to open neighboring RyRs. See Figure S3.**

| Condition | [Ca] at the nearest closed RyR, μM | [Ca] at RyR#2, μM |
|---|---|---|
| Open RyR#1 | 4.904 | 8.234 |
| Open RyR#2 | 8.66 | 22.256 |
| Open RyR#1 and RyR#2 (independent) | 13.564 | 30.49 |
| Open RyR#1 and RyR#2 (real) | 13.094 | 28.701 |
| Difference of independent and real | 0.47 | 1.789 |
| relative change in % | 3.465 | 5.867 |

**Table S3 - Supplemental calculations & results (median extinction times) for the main text Figure 2.**

Pairs of $\gamma$ and $\lambda$ values (highlighted) were constructed to have $h=0$ in a CRU with 9x9 RyRs. This is a copy of Excel spreadsheet; each column is explained below:
$C$: closing rate $C=0.117$ ms$^{-1}$= const.
$CaJSR$: $Ca_{jSR}=100$ µM=const.
$\psi(U)$: the value of interaction profile at the distance of the nearest RyR neighbor; $\psi(r)$ is given in Figure S2A, i.e. we take here the $\psi$ value at $r=1$ for the 0.1 mM $Ca_{JSR}$ curve (red line).
**Int0.5toInf** = $\int_{r>.5} \psi(r)dr$. The integral is calculated for interaction profile $\psi(r)$, that is the red line in Figure S2A. The integral is taken from $r=0.5$ to infinity (to the end of the grid, that is 4.33 in our case).

**SpaceInt**=$2*\pi*int0.5toInf/\psi(U)$. This is the normalized 2d integral in Equation [10] of the supplement.
$\gamma$: independent variable here, it varies from 0.02 to 0.15 1/µM.
$\beta=\psi(U)*\gamma/4$, i.e. Equation 9b in the supplement.
$\lambda=C*exp(-2*\beta*SpaceInt)$. This is the solution of the Equation [10] in the supplement for $h=0$. Each $\lambda$ value was calculated for each $\gamma$ (independent variable).

$$h = \frac{1}{2\beta}\ln\left(\frac{\lambda}{C}\right) + 2\pi\int_{r>.5}\phi(r,Ca_{JSR})dr = \frac{1}{2\beta}\ln\left(\frac{\lambda}{C}\right) + SpaceInt = 0$$

**extin time**: Median extinction time for 100 sparks simulated with parameters in each row.
$h=ln(\lambda/C)/(2*\beta)+SpaceInt$: analog of magnetic field that must be 0. We calculated $h$ just to make sure that it is indeed 0.

| C | CaJSR | $\psi(U)$ | int0.5toInf | SpaceInt | $\gamma$ | $\beta$ | $\lambda$ | extin time | h |
|---|---|---|---|---|---|---|---|---|---|
| 1/ms | mkM | mkM | mkM | NoDim | 1/mkM | NoDim | ms-1 | ms | NoDim |
| 0.117 | 100 | 7.47087 | 15.1703 | 12.7586 | 0.02 | 0.0373544 | 4.51E-02 | >10000 | 0 |
| 0.117 | 100 | 7.47087 | 15.1703 | 12.7586 | 0.03 | 0.0560315 | 2.80E-02 | >10000 | 0 |
| 0.117 | 100 | 7.47087 | 15.1703 | 12.7586 | 0.04 | 0.0747087 | 1.74E-02 | >10000 | 0 |
| 0.117 | 100 | 7.47087 | 15.1703 | 12.7586 | 0.05 | 0.0933859 | 1.08E-02 | >10000 | 0 |
| 0.117 | 100 | 7.47087 | 15.1703 | 12.7586 | 0.053 | 0.098989 | 9.36E-03 | >10000 | 0 |
| 0.117 | 100 | 7.47087 | 15.1703 | 12.7586 | 0.055 | 0.1027245 | 8.51E-03 | 7397.46 | 0 |
| 0.117 | 100 | 7.47087 | 15.1703 | 12.7586 | 0.058 | 0.1083276 | 7.37E-03 | 2473.98 | 0 |
| 0.117 | 100 | 7.47087 | 15.1703 | 12.7586 | 0.06 | 0.1120631 | 6.70E-03 | 1110.41 | 0 |
| 0.117 | 100 | 7.47087 | 15.1703 | 12.7586 | 0.07 | 0.1307402 | 4.16E-03 | 188.308 | 0 |
| 0.117 | 100 | 7.47087 | 15.1703 | 12.7586 | 0.08 | 0.1494174 | 2.58E-03 | 97.4777 | 0 |
| 0.117 | 100 | 7.47087 | 15.1703 | 12.7586 | 0.09 | 0.1680946 | 1.60E-03 | 78.6526 | 0 |
| 0.117 | 100 | 7.47087 | 15.1703 | 12.7586 | 0.1 | 0.1867718 | 9.96E-04 | 69.4714 | 0 |
| 0.117 | 100 | 7.47087 | 15.1703 | 12.7586 | 0.11 | 0.2054489 | 6.19E-04 | no data | 0 |
| 0.117 | 100 | 7.47087 | 15.1703 | 12.7586 | 0.12 | 0.2241261 | 3.84E-04 | no data | 0 |
| 0.117 | 100 | 7.47087 | 15.1703 | 12.7586 | 0.13 | 0.2428033 | 2.38E-04 | no data | 0 |
| 0.117 | 100 | 7.47087 | 15.1703 | 12.7586 | 0.14 | 0.2614805 | 1.48E-04 | no data | 0 |
| 0.117 | 100 | 7.47087 | 15.1703 | 12.7586 | 0.15 | 0.2801576 | 9.19E-05 | 56.0243 | 0 |

**Table S4 – Values of *h* for the results in Figure 3 (main text).**

| JSR [Ca], mM | h |
|---|---|
| 0.0125 | -100 |
| 0.025 | -44 |
| 0.0375 | -25 |
| 0.044 | -20 |
| 0.05 | -16 |
| 0.0625 | -10 |
| 0.075 | -7 |
| 0.1 | -2 |

**Supporting Material Reference List**

1. Maltsev, A.V., V.A. Maltsev, and M.D. Stern. 2017. Clusters of Ca release channels harness the Ising phase transition to confine their elementary intracellular signals. Proc Natl Acad Sci U S A 114:7525–7530.
2. Levin, D.A., Y. Peres, and E.L. Wilmer. 2008. Markov Chains and Mixing Times. American Mathematical Society.
3. Martinelli, F. 1999. Lectures on Glauber Dynamics for Discrete Spin Models. *In* Lectures on Probability Theory and Statistics. Bernard P, editor. Springer-Verlag, Berlin Heidelberg.
4. Evans, M. Lecture Notes. Section 11: Exact results for the Ising Model https://www2.ph.ed.ac.uk/~mevans/sp/sp10.pdf.
5. Baxter, R.J. 1989. Exactly Solved Models in Statistical Mechanics. Academic Press.
6. Stern, M.D., E. Rios, and V.A. Maltsev. 2013. Life and death of a cardiac Ca spark. J Gen Physiol 142:257-274.
7. Laver, D.R., C.H. Kong, M.S. Imtiaz, and M.B. Cannell. 2013. Termination of Ca-induced Ca release by induction decay: an emergent property of stochastic channel gating and molecular scale architecture. J Mol Cell Cardiol 54:98-100.

# Movies y kj 'tj gk 'y gb links and regends

Online Movie S1
https://drive.google.com/open?id=1lSY-UQAzWBsTk41xsVMOmzPp3QJZTe-u
Supplements Figure 4A. Normal spark termination in supercritical regime. Total simulation time is 80 ms. $Ca_{FSR}$=1 mM, $\tau_{fill}$=11ms.

Online Movie S2
https://drive.google.com/open?id=1FCiGg-Wv-4ede7DtD568DRN8-aC7cTNW
Supplements Figure 4B. Pathological spark with synchronized leak via JSR quick refiling with Ca in supercritical regime. Total simulation time is 80 ms. $Ca_{FSR}$=1 mM, $\tau_{fill}$=0.5ms.

Online Movie S3
https://drive.google.com/open?id=1-RFN4AdgUsRCLlqlI_1g67n1wPMuZ_an
Supplements Figure 4C. Pathological spark with synchronized leak under Ca overload in supercritical regime. Total simulation time is 80 ms. $Ca_{FSR}$=5 mM, $\tau_{fill}$=11 ms.

Online Movie S4
https://drive.google.com/open?id=1Aq7t3AFpe6PSZ1Yqd3LufNH2ScatR2GD
Supplements Figure 5B. Disorder leak that oscillates near criticality ($\beta^*$). Total simulation time in the movie is 500 ms. $\lambda = 10*\lambda_{fit}$.

Online Movie S5
https://drive.google.com/open?id=1SArmee48phh6wBVMufiO2wlh5VBRwLlP
Supplements Figure 5C. Persistent disorder leak in subcritical regime. Total simulation time in the movie is 200 ms. $\lambda = 100*\lambda_{fit}$.

Online Movie S6
https://drive.google.com/open?id=18mqkf9hje6wHxM0aLDAEO5ywhNjagVqr
Supplements Figure 6. A smaller Peierl's contour (white lines) in case of supercritical, synchronized leak. Total simulation time is 200 ms. $Ca_{FSR}$=1 mM, $\tau_{fill}$=0.5ms (similar to Movie S2).

Online Movie S7
https://drive.google.com/open?id=1dBRlV5vuuj4vIOllJeGRW8qLvWPk3aGv
Supplements Figure 6. A larger Peierl's contour (white lines) in case of subcritical disorder leak. Total simulation time is 200 ms. $\lambda = 100*\lambda_{fit}$.